\documentclass[sigconf]{acmart}
\renewcommand\footnotetextcopyrightpermission[1]{}
\usepackage{aligned-overset}
\usepackage{hyperref}
\usepackage{graphicx}
\usepackage{makecell}
\usepackage{multirow}
\usepackage{tabularx}
\usepackage{multirow}
\usepackage{tabularx}
\usepackage{enumerate}
\usepackage{bbm}
\usepackage{latexsym}
\usepackage{wasysym}

\newtheorem{theorem}{\textbf{Theorem}}

\begin{document}
\fancyhead{}

\title{Understanding the Identity-Transformation Approach in OIDC-Compatible Privacy-Preserving SSO Services}

\author{
{\rm Jingqiang Lin$^{\ddag}$, Baitao Zhang$^{\ddag}$, Quanwei Cai$^{\P}$, Wei Wang$^{\ddag}$, Jiwu Jing$^{\diamondsuit}$, and Huiyang He$^{\ddag}$}\\
$\ddag$ School of Cyber Security, University of Science and Technology of China\\
$\P$ Beijing Zitiao Network Technology Co., Ltd, China\\
${\diamondsuit}$ School of Cryptology, University of Chinese Academy of Sciences}

\begin{abstract}
Single sign-on (SSO) enables a user to log into multiple websites, called \emph{relying parties} (RPs),
            by her username and credential set up in another trusted web system, called the \emph{identity provider} (IdP).
\emph{Identity transformations} are proposed in UppreSSO to provide privacy-preserving SSO services, preventing both IdP-based login tracing and RP-based identity linkage.
While the security and privacy guarantees of UppreSSO have been proved,
    several essential issues on the identity-transformation approach are not well studied.
In this paper,
    we comprehensively investigate this approach as below.
Firstly, several suggestions
             to efficiently integrate identity transformations into OpenID Connect (OIDC) are explained.
Then, 
    we uncover the relationship between identity transformations in SSO
            and oblivious pseudo-random functions (OPRFs),
    and present two variations of the properties required for SSO security
            as well as other requirements,
     to analyze existing OPRF protocols.
Finally, new identity transformations different from those proposed in UppreSSO,
    are constructed based on some OPRFs.
To the best of our knowledge,
    this is the first time to uncover the relationship between identity transformations in SSO services and OPRFs,
        and prove the SSO-related properties (i.e., output uniqueness, key-identifier freeness, and collision resistance on 1st/2nd-input) of typical OPRFs.
  %      in addition to the basic properties of correctness, obliviousness and pseudo-randomness.
\end{abstract}

\maketitle

%\begin{IEEEkeywords}
%Single sign-on; privacy; security.
%\end{IEEEkeywords}
%
%\newtheorem{lemma}{Lemma}
%\newtheorem{assumption}{Assumption}
%\newtheorem{tss-bqs-def}{Definition}
%\newtheorem{theorem}{Theorem}[section]

\hyphenation{time-stamp}

\section{Introduction}
\label{s1:introduction}
% SSO用户隐私很重要，但是做到平衡：隐私、兼容、安全不容易。
Single sign-on (SSO) \cite{rfc6749,SAMLIdentifier,OpenIDConnect} 
        facilitates a user to visit multiple websites, called \emph{relying parties} (RP),
            by her username and credential set up in another trusted web system, called the \emph{identity provider} (IdP).
OpenID Connect (OIDC) \cite{OpenIDConnect} is the most popular SSO protocol in the Internet, as users access the services from a commercial-off-the-shelf (COTS) browser.

SSO enables an RP to delegate user authentication to the IdP, which issues an \emph{identity token} for an authenticated user to visit the RP,
    but the original designs of OIDC raise the concerns on user privacy.
For instance, when requesting identity tokens from an IdP to visit an RP,
    a user submits the target RP's identity,
    so that a curious IdP could trace the user's login activities \cite{BrowserID,SPRESSO,uppresso-arxiv-all},
    called \emph{IdP-based login tracing}.
Meanwhile,
    if a user visits different RPs with tokens enclosing an identical user identity,
        colluding RPs could exploit this identity to link the accounts across these RPs to profile the user \cite{maler2008venn,uppresso-arxiv-all,NIST2017draft},
called \emph{RP-based identity linkage}.

These privacy threats result from the common requirements of SSO, but not the designs of specific protocols,
    because an identity token needs to enclose the identities of  the target RP and the authenticated user \cite{rfc6749,SAMLIdentifier,OpenIDConnect},
    to (\emph{a}) prevent malicious RPs from replaying received identity tokens to some honest RP as victim users and (\emph{b}) identify the initiating user at the visited RP.

Several solutions provide OIDC-compatible SSO services \cite{BrowserID,SPRESSO,NIST2017draft,uppresso-arxiv-all,POIDC,save-flow,ARPSSO,up-sso,miso},
    while preventing user privacy leakages (i.e., IdP-based login tracing, RP-based identity linkage, or both).
However,
    none of them except UppreSSO \cite{uppresso-arxiv-all} offers all describes features.
Some schemes prevent only one privacy threat \cite{BrowserID,SPRESSO,NIST2017draft,POIDC,save-flow},
    and the others introduce extra trusted servers (or components) in addition to the honest IdP \cite{miso,up-sso,SPRESSO,ARPSSO} or impractical computation costs \cite{POIDC}.
UppreSSO proposes an identity-transformation approach to protect users against two privacy risks,
            without introducing trusted servers more than an honest-but-curious IdP.
In particular, when a user visits an RP in UppreSSO,
    \emph{ephemeral pseudo-identities} are generated for the visited RP and the authenticated user,
    and signed in an identity token by the IdP,
which allows the user to log into the target RP as her \emph{permanent account}.

While the identity-transformation approach has been described,
         proved, implemented, and evaluated in the implicit flow of OIDC \cite{uppresso-arxiv-all}, 
        and integrated in the authorization code flow \cite{ARPSSO},
   there are still essential issues not well studied.
Firstly, 
    the integration of identity transformations in SSO services sometimes
    introduces unnecessary calculations \cite{uppresso-arxiv-all,ARPSSO} or extra trusted servers \cite{ARPSSO}. % and even privacy leakage \cite{ARPSSO}.
In this paper we explicitly present suggestions for more efficient integrations.
Moreover, the suggestion for an RP to accept identity tokens efficiently, inspires us to analyze variations of the properties required for secure SSO,
       i.e., user identification and RP designation,
            either \emph{with} or \emph{without} checking the RP's pseudo-identity in received tokens.
Secondly, %as ArpSSO \cite{ARPSSO} shifts the identity transformations on elliptic curves \cite{uppresso-arxiv,uppresso-conference} to a finite field,
    we try to construct more identity transformations qualified for privacy-preserving SSO (i.e., design a generalized UppreSSO system).
Then, the relationship between the identity transformations
            and oblivious pseudo-random functions (OPRFs) is uncovered,
which helps us to analyze existing OPRF schemes.
 % following the variations of user identification and RP designation as well as other properties required for SSO.
Finally, based on the analysis results,
    new identity transformations 
    are constructed for generalized UppreSSO systems, satisfying different variations of security requirements as well as other properties.

To the best of our knowledge,
    this is the first time to (\emph{a}) uncover the relationship between identity transformations in privacy-preserving SSO services and general OPRFs,
        and (\emph{b}) prove the SSO-related properties of typical OPRFs (i.e., output uniqueness, key-identifier freeness and collision resistance on 1st/2nd-input),
        which are very different from the basic OPRF properties.
         % of correctness, obliviousness and pseudo-randomness.

The remainder of this paper is organized as below.
Section \ref{IDT-uppresso} describes the identity-transformation approach.
Section \ref{compare-sso} discusses the integration of identity transformations in OIDC.
We demonstrate new identity transformations in Section \ref{idt-oprf}, and discuss related work in Section \ref{related-work}.
Section \ref{conclusion} concludes this paper.

\section{Identity Transformations in UppreSSO}
\label{IDT-uppresso}
This section describes the identity-transformation approach proposed in UppreSSO \cite{uppresso-arxiv-all}
 for privacy-preserving SSO services.
Only the fundamental designs of this approach are included in this section,
    and more details can be found in \cite{uppresso-arxiv-all}.

\subsection{System Model and Initialization}
\label{system-model}
An SSO system consists of 
    $p$ RPs, $s$ users, and an \emph{honest-but-curious} IdP signing identity tokens for a user to visit these RPs.
\emph{Malicious} adversaries could fully control some \emph{RPs} and \emph{users},
    attempting to break the security or privacy guarantees for honest RPs and users.

Authenticated and confidential links are assumed between honest entities, 
 and the adopted cryptographic primitives are secure. The software stack of an honest entity is correctly implemented to deliver messages to receivers as expected.
That is,
    authenticity, confidentiality and integrity of an identity token 
are ensured by secure communications in the system \cite{uppresso-arxiv-all}.
%(i.e., HTTPS among entities and the security mechanism of \verb+postMessage+ in browsers \cite{GoogleIdIntegrate,uppresso-arxiv,uppresso-conference,dimvaLiM16}) and digital signatures (i.e., signed identity tokens and RP certificates \cite{uppresso-arxiv,uppresso-conference,ARPSSO}) in the system.
Meanwhile,
    there is no direct privacy leakage in the requesting, receiving and forwarding of identity tokens. %  due to the adopted web mechanisms \cite{uppresso-conference}.

In their registrations, 
unique $ID_U$ and $ID_{RP}$ are assigned to a user and an RP by the honest IdP, respectively.
Then, a registered user with $ID_U$ is automatically assigned an account $Acct = \mathcal{F}_{{Acct\ast}}(ID_U, ID_{RP})$ at the RP with $ID_{RP}$.
An account is \emph{meaningful} at an RP if it belongs to some \emph{registered} user at this RP,
while a \emph{meaningless} account $A_m$ at the $j$-th RP does not belong to any registered user, i.e., $\forall ID_{U_i} \in \mathbbm{ID}_U = \{ID_{U_{i; 1 \leq i \leq s}}\}: A_m \neq \mathcal{F}_{Acct\ast}(ID_{U_i}, ID_{RP_j})$.

%Every RP synchronizes all accounts at it from the honest IdP.
%An account belonging to some registered user is \emph{meaningful},
%while \emph{meaningless} accounts at an RP do not belong to any registered user.

\subsection{The SSO Login Flow with Identity Transformations}
\label{imp-flow-idt}
The identity transformations of UppreSSO \cite{uppresso-arxiv-all} are proposed on elliptic curves.
$\mathbb{E}$ is an elliptic curve over a finite field $\mathbbm{F}_q$, where $G$ is a generator on $\mathbb{E}$ and the order of $G$ is a prime $n$.

The IdP randomly selects $ID_U = u \in \mathbbm{Z}_n$ for each registered user,
    while it randomly selects $r \in \mathbbm{Z}_n$ and assigns $ID_{RP} = [r]G$ to an RP,
    where $[r]G$ denotes the addition of $G$ on the elliptic curve $r$ times.
Then, $Acct = \mathcal{F}_{{Acct\ast}}(ID_U, ID_{RP}) = [ID_U]ID_{RP} = [ur]G$.

$ID_{RP}$ and $Acct$ are \emph{publicly known},
    but $u$ and $r$ are kept \emph{secret} to all entities except the IdP.
%$ID_U$ is processed only by the IdP internally and never enclosed in tokens or any messages,
%    and $r$ is not used any more after $ID_{RP}$ is calculated in the registration.
Each RP locally maintains its meaningful account set \cite{uppresso-arxiv-all},
i.e., 
$\{Acct_{i; 1 \leq i \leq s}\}_j=\{[u_{i; 1 \leq i \leq s} r_j]G\}$, for all registered users.
 Section \ref{account-syn} discusses account synchronization.

In addition to $Acct = \mathcal{F}_{{Acct\ast}}()$,
    the following functions are designed in UppreSSO \cite{uppresso-arxiv-all}:
\begin{itemize}
  \item $PID_{RP} = \mathcal{F}_{PID_{RP}}(ID_{RP}, t) = [t]ID_{RP} = [tr]G$, where $t$ is a random number in $\mathbbm{Z}_n$.
  \item $PID_U = \mathcal{F}_{PID_U}(ID_U, PID_{RP}) = [ID_U]PID_{RP} = [utr]G$.
  \item $Acct = \mathcal{F}_{Acct}(PID_U, t) = [t^{-1}]PID_{U} = [t^{-1}utr]G = [ur]G = \mathcal{F}_{Acct\ast}(ID_U, ID_{RP})$.
\end{itemize}

While $\mathcal{F}_{Acct\ast}()$ determines a registered user's account unique at every RP,
    in a login $\mathcal{F}_{PID_{RP}}()$ and $\mathcal{F}_{PID_U}()$ transform 
        an RP's identity and a user's identity into two \emph{pseudo-identities} to prevent IdP-based login tracing and RP-based identity linkage, respectively.
Finally, $\mathcal{F}_{Acct}()$ enables an RP to derive a user's \emph{permanent} account based on \emph{ephemeral} $PID_{U}$,
    equal to the one determined by $\mathcal{F}_{Acct\ast}()$.

The SSO login flow with identity transformations works as follows \cite{uppresso-arxiv-all}.
\begin{enumerate}
  \item When attempting to access protected resources at an RP,
   a user obtains $ID_{RP}$ of her target RP, randomly selects $t$, and calculates $PID_{RP} = \mathcal{F}_{PID_{RP}}(ID_{RP}, t)$. Then, she requests a token for $PID_{RP}$ from the IdP. %and $t$ is sent to the RP.
  \item The IdP authenticates the user, if not authenticated yet.
  It calculates $PID_U = \mathcal{F}_{PID_U}(ID_U, PID_{RP})$ where $ID_U$ is the authenticated user's identity, signs an identity token binding $PID_{RP}$ and $PID_U$, and returns this token to the user.
  \item The initiating user forwards the signed token to the visited RP along with $t$. The RP verifies the received token, extracts $PID_U$ from it, calculates $Acct = \mathcal{F}_{Acct}(PID_U, t)$, and allows the user to log in as $Acct$
      if it is a meaningful account (i.e., in its account set).
\end{enumerate}

We particularly note two issues \cite{uppresso-arxiv-all}.
%In UppreSSO user operations are implemented on a COTS browser with scripts, 
%    and it works compatibly with OIDC services.
%The trusted part of user-agent scripts is downloaded from the honest-but-curious IdP,
%        and the other script is from the visited RP to forward identity tokens.
%During the login,
%     $t$ is kept \emph{secret} to the honest IdP, known to only the initiating user and the visited RP.
%Moreover, $t$ is \emph{ephemeral} so that the services are accessed from COTS browsers;
%otherwise, a long-term user secret requires a browser extension \cite{uppresso-conference}.
%
First of all, $PID_{RP}$ is calculated based on $ID_{RP}$ of the visited RP (but not any other RP) by the user,
    so that the target RP is correctly designated.
Secondly, after verifying the IdP's signature on a received token, an RP does \emph{not} check whether $PID_{RP}$ enclosed in the token is equal to $[t]ID_{RP}$ or not.
    This design does not lead to attacks, because an unmatching token results in meaningless accounts and will be rejected \cite{uppresso-arxiv-all}.

%Finally, the calculation of $PID_{RP}$ by the user (but not the RP) before an identity token is requested,
%    results in the correctly designated RP by the token (see Section \ref{uppresso-security} for details), 
%    while $PID_{RP}$ checking by an RP before the RP accepts a signed token, is eliminated by the properties of the proposed identity transformations in UppreSSO.

\subsection{Security and Privacy Properties of UppreSSO}
\label{SP-ID-Proof}

\subsubsection{Security}
\label{uppresso-security}
In UppreSSO with identity transformations, an identity token denoted as $TK$,
    is signed by the IdP to bind $PID_{RP}$ and $PID_U$,
where (\emph{a}) the RP with $ID_{RP}$ is designated by $TK$ as $PID_{RP} = [t]ID_{RP}$ is calculated
    and (\emph{b}) $PID_U = [ID_U]PID_{RP}$ is calculated based on the authenticated user's identity $ID_U$.

As authenticity, confidentiality and integrity of an identity token 
have been ensured,
     the \emph{sufficient} conditions of secure SSO services are listed as below \cite{uppresso-arxiv-all},
    for any signed token $TK$.

\noindent \textbf{User Identification:} %At the designated RP, $TK$ identifies only the authenticated user who requests this token from the IdP. That is, 
\emph{Based on $TK$ the designated honest RP derives only the meaningful account belonging to the user requesting $TK$.}

\noindent \textbf{RP Designation:} %Based on $TK$, only the designated (honest) RP derives meaningful accounts belonging to registered users. That is, 
\emph{At any honest RPs other than the designated one, no meaningful account is derived based on $TK$.}

The four functions proposed in Section \ref{imp-flow-idt} are proved to satisfy user identification and RP designation \cite{uppresso-arxiv-all}
with an honest IdP but against malicious RPs and users. % even if an (honest) RP accepts any verified tokens \emph{without} checking $PID_{RP}$.
%We will analyze these properties in Section \ref{idt-oprf} with more extensions.

These security properties prevent tokens from being exploited by adversaries (i.e., malicious RPs and users),
if \emph{the target RP of an identity token is designated by the initiating user who requests this token}:
    She obtains the target RP's identity (i.e., $ID_{RP}$) and calculates $PID_{RP}$ based on it, as mentioned in Section \ref{imp-flow-idt}.

We elaborate ``the designated RP'' of $TK$,
        by exhibiting an attack due to vulnerable $PID_{RP}$ calculations.
If an RP selects $t$, calculates $PID_{RP}=[t]ID_{RP}$, and sends $PID_{RP}$ to a user who directly requests a token for $PID_{RP}$, 
impersonation attacks happen as below, 
    even when an (honest) RP additionally checks $PID_{RP}$ in a received token.
For example,
        a malicious user receives $PID_{RP}$ when visiting an honest RP.
This malicious user does not request identity tokens for $PID_{RP}$,
    but forwards $PID_{RP}$ to some colluding RP.
Once a user visits this colluding RP,
        it sends $PID_{RP}$ to this victim user to obtain a token,
        and shares it with colluding adversaries.
Finally,
the malicious user could exploit this token binding $PID_{RP}$,
            to successfully log into the \emph{honest} RP as the victim's account.

The above attack results from the fact that the colluding RP is \emph{not} designated in the signed token, because $PID_{RP}$ is not calculated based on its RP identity.
A naive prevention against such impersonations is
        to \emph{additionally} check $PID_{RP} = [t]ID_{RP}$ by the user who receives $PID_{RP}$ and also $t$ from the visited RP,
            before requesting a token.
Thus, $PID_{RP}$ is calculated \emph{repeatedly}:
    The RP selects $t$, calculates $PID_{RP} = [t]ID_{RP}$,
        and sends them to the user, who checks whether $PID_{RP}$ is equal to $[t]ID_{RP}$ by calculating $[t]ID_{RP}$ again.\footnote{This inefficient design was originally presented in the preliminary version of UppreSSO \cite{uppresso-arxiv-all}, and then inherited by ArpSSO \cite{ARPSSO} in 2024.
        It is improved later \cite{uppresso-arxiv-all}, and this repeated calculation of $PID_{RP}$ is finally eliminated in the integration of identity transformations.}

On the contrary, the calculation of $PID_{RP}=[t]ID_{RP}$ by users % (more exactly, the honest user-i script in the user browser) \cite{uppresso-conference},
    prevents such impersonation attacks effectively and efficiently,
 because the RP with $ID_{RP}$ is designated 
 as $PID_{RP}=[t]ID_{RP}$ is calculated by the user who requests tokens from the IdP.

\subsubsection{Privacy}
\label{uppresso-privacy}
%As there is no direct privacy leakage in the requesting, receiving and forwarding of identity tokens \cite{uppresso-conference},
% We analyze the privacy properties of UppreSSO as below.
An IdP traces a user's login activities by the RP (pseudo-)identities in identity-token requests,
    and 
IdP untraceability is analyzed against the honest-but-curious IdP.

\noindent \textbf{IdP Untraceability:} 
%\emph{The IdP cannot distinguish $PID_{RP}$ from a uniformly random variable. That is, 
\emph{The honest-but-curious IdP learns nothing on the visited RP from the identity-token requests.}

In UppreSSO among the messages received by the IdP,
    only $PID_{RP}$ is calculated based on $ID_{RP}$.
The IdP cannot distinguish $PID_{RP}$ from a uniformly random variable,
    which has been proved \cite{uppresso-arxiv-all},
    and then IdP untraceability is ensured in UppreSSO.

RP-based identity linkage is analyzed against a group of semi-honest RPs and users \cite{uppresso-arxiv-all}.
These adversaries do not actively violate the protocol
    but share their information, attempting to link an honest user's login activities.
For example, when an honest user visits a semi-honest RP,
    adversaries attempt to link this login to some login visiting another RP in the group
    (i.e., link two accounts across these RPs).

\noindent \textbf{RP Unlinkability:}
\emph{Semi-honest RPs and users
 cannot collude to link any login initiated by an honest user visiting a semi-honest RP, to any subset of logins visiting any other colluding RPs by honest users.}

All the information that an RP learns in a login is denoted as a \emph{r-tuple} $(ID_{RP}, t, Acct)=([r]G, t, [ur]G)$,
    and RP unlinkability is proved in UppreSSO \cite{uppresso-arxiv-all}.

%\section{OIDC-Compatible Privacy-Preserving SSO}
\section{OIDC-Compatible Privacy-Preserving SSO Services}
\label{compare-sso}
We explain the SSO services specified by OIDC,
 and investigate the integration of identity transformations in OIDC.

\subsection{OIDC-Compatible SSO}
\label{comparison}

OIDC \cite{OpenIDConnect} is the most popular SSO protocol.
It specifies endpoints (i.e., the web methods accessed through HTTP) for the IdP and RPs,
    so that a user requests, receives and forwards identity tokens from a COTS browser.
The IdP's authorization endpoint defines the interface to issue tokens in the implicit login flow of OIDC and to issue authorization codes in the authorization code flow,
    while the interface to receive tokens is defined in the RP's redirection endpoint.
The authorization code flow additionally requires the IdP's token endpoint for an RP to retrieve tokens in the back channels.

OIDC-compatible UXs are designed to provide different user experiences
    by leveraging different web pages and scripts,
        but the endpoint interfaces are kept unchanged.
For example,
    pop-up UX provides an in-context user experience \cite{dimvaLiM16,GoogleIdIntegrate,uber},
    while redirect UX \cite{OpenIDConnect,rfc6749,GoogleIdIntegrate}
            works with only one browser window,
            either in the implicit flow or the authorization code flow of OIDC.

While the original OIDC protocol does not prevent  IdP-based login tracing or RP-based identity
linkage,
various privacy-preserving SSO services \cite{BrowserID,uppresso-arxiv-all,SPRESSO,NIST2017draft,POIDC,save-flow,up-sso,miso} are implemented compatibly with OIDC, by using the specified endpoint interfaces.

\subsection{Integrating Identity Transformations into OIDC Compatibly}
In UppreSSO user operations are conducted in browser
by web pages and scripts as in other OIDC-compatible SSO services,
    and the other operations are done by the IdP web server and RP servers which follow the endpoint interfaces specified in OIDC.

As the IdP receives $PID_{RP}$ (but not $ID_{RP}$) in an identity-token request,
    it cannot inform the browser about the target RP's redirection endpoint URL and
        the token binding $PID_{RP}$ is issued to an ``empty'' URL.
An RP certificate signed by the IdP during the RP's initial registration,
    is introduced to bind $ID_{RP}$ and the RP's endpoint URL.
When a user clicks the RP's SSO login button in her browser,
        the RP certificate is carried to an IdP web page as a fragment and verified by the user-i script which is downloaded in this page.
The user-i script extracts $ID_{RP}$, calculates $PID_{RP} = [t]ID_{RP}$,
        and requests an identity token from the IdP web server.
Then, the user-i script forwards the token as well as $t$ to the RP,
    which is restricted by the origin of the URL extracted from RP certificates.

Besides, there are several web page settings to prevent direct privacy information in the requesting of identity tokens \cite{uppresso-arxiv-all},
    e.g., an HTTP \texttt{referer} header about the RP web server.

\subsection{Efficient Integrations of Identity Transformations}
\label{arpsso-improvement}
ArpSSO \cite{ARPSSO} integrated the identity transformations of UppreSSO in the authorization code flow of OIDC.\footnote{In ArpSSO \cite{ARPSSO} the identity transformations of $ID_{RP}$-$PID_{RP}$ and $ID_U$-$PID_U$ are called \emph{RP anonymization} and \emph{user identity mix-up}, respectively.}
%IdP-based login tracing and RP-based identity linkage are prevented in ArpSSO by the integrated identity transformations.
By demonstrating the improvements on ArpSSO,
we propose several suggestions for more efficient integrations of the identity transformations in OIDC, 
some of which have been proposed but not clearly explained.\footnote{Some suggestions are also applicable to BISON \cite{bison} and OPPID \cite{OPPID}.}

First of all, ArpSSO introduces extra trusted servers to distribute and verify the web scripts,
    which are unnecessary due to the common honest-but-curious assumptions of the IdP in ArpSSO,
    UppreSSO and other privacy-preserving SSO systems.
Only the user-i script is assumed to be honest:
    As it is downloaded from the honest IdP web server over HTTPS/TLS, this assumption holds without extra trusted servers.
%On receiving an identity-token request, the IdP web server checks the included HTTP \texttt{origin} header to ensure it is sent by the user-i script,
%which will honestly verify $Cert_{RP}$ to ensure the correct relationship between $ReEnptUrl_{RP}$ and $ID_{RP}$.
Note that the user-r script downloaded from an RP might be malicious,
 which is tolerated \cite{uppresso-arxiv-all,ARPSSO}.

Secondly, while unnecessity of $PID_{RP}$ calculations by an RP,
    either before a token is requested
        or after the token is signed,
is proved in the implicit flow of OIDC \cite{uppresso-arxiv-all},
    it is also applicable to the integration in the authorization code flow.
We explain this by improving ArpSSO as follows.

ArpSSO shifts the identity transformations of UppreSSO on elliptic curves into a finite field $\mathbbm{F}_q$,
where $ID_U = u$, $ID_{RP} = g^r$, $PID_{RP} = ID_{RP}^{t_1} = g^{rt_1}$ and $PID_U = PID_{RP}^{u}=g^{rut_1}$.
Meanwhile, it employs PS signing \cite{PS-sig} to bind $ID_{RP}$ and an RP's domain:
In the registration an RP obtains a PS signature $(g^r, g^{r(x+y_1d)})$ for its domain $d$, where $(x,y_1)$ is the IdP's PS signing key.
This facilitates users to calculate $PID_{RP}$ based on the visited RP's identity.

Let $\sigma_2 = g^{r(x+y_1d)}$,
    and the login flow in ArpSSO is described as below.
It is worth nothing that,
    $t_1$ and $t_2$ in this description, correspond to $k$ and $t$ in the original expressions \cite{ARPSSO}, respectively.
The random number $t_2$, is generated for RP anonymous authentication \cite{ELPASSO} to the IdP in the retrieval of tokens in the authorization code flow, unrelated to the identity transformations.
\begin{enumerate}
  \item  When visited by a user, an RP randomly selects $t_1, t_2 \in \mathbbm{F}_q$, and calculates $(g^{rt_1}, (\sigma_2 g^{rt_2})^{t_1})$ which is sent to the user along with $t_2$. Moreover, $(g^{rt_1}, (\sigma_2 g^{rt_2})^{t_1})$ is kept by the RP for the subsequent anonymous authentication to the IdP.
  \item  The user calculates $v = (1/g^{rt_1})^{t_2}$ and $w = v(\sigma_2 g^{rt_2})^{t_1}$,
   and verifies whether $(g^{rt_1}, w)$ is a valid PS signature for $d$, which is the target RP's domain.
  \item  The user requests an identity token for $PID_{RP} = g^{rt_1}$,
    and receives an authorization code from the IdP.
This code is forwarded to the RP,
        and used to retrieve the token binding $PID_{RP} = g^{rt_1}$ and $PID_U = PID_{RP}^u = g^{urt_1}$.
The RP finally allows the user to log in as $Acct = (g^{urt_1})^{1/t_1} = g^{ur}$.
\end{enumerate}

In the above procedure, $PID_{RP} = ID_{RP}^{t_1} = g^{rt_1}$ is first calculated by the visited RP
 and later checked actually by the user calculating $v = (1/g^{rt_1})^{t_2}$ and $w = v(\sigma_2 g^{rt_2})^{t_1} = \sigma_2^{t_1}$.
We improve it to eliminate the repeated calculation of $PID_{RP}$ as follows.
\begin{enumerate}
  \item   The RP sends $(g^r, \sigma_2)$ to a user visiting protected resources.
  After verifying that $(g^r, \sigma_2)$ is a valid PS signature for $d$,
   the initiating user randomly selects $t_1$, calculates $g^{rt_1}$, and requests an identity token for $g^{rt_1}$.
  \item   An identity token binding $PID_{RP} = g^{rt_1}$ and $PID_U = g^{urt_1}$ is signed by the IdP. %This token allows the user to log in.
  \item The user sends $g^{rt_1}$ and $t_1$ to the RP.
The RP randomly selects $t_2$ and calculates only $(\sigma_2 g^{rt_2})^{t_1}$ to construct $(g^{rt_1}, (\sigma_2 g^{rt_2})^{t_1})$ for the subsequent anonymous authentication to the IdP.
\end{enumerate}

Thus, $PID_{RP} = g^{rt_1}$ is calculated \emph{only} by the user.
We eliminate the calculations of $v = (1/g^{rt_1})^{t_2}$ and $w = v(\sigma_2 g^{rt_2})^{t_1}$,
    while the security and privacy guarantees are still strictly provided.

Thirdly, to efficiently integrate identity transformations in the authorization code flow of OIDC, in the retrieval of tokens
    we recommend (\emph{a}) the widely-used OIDC option of proof key for code exchange (PKCE) \cite{pkce-oauth}, and (\emph{b})
    anonymous credentials 
      such as ring signature \cite{ring-sig} and privacy-pass token \cite{privacypass,trusttoken},
     to serve for RP anonymous authentication to the IdP.
In every login the target RP generates a PKCE verifier and hashes it into a challenge.
        This PKCE challenge is sent to the user, and forwarded to the IdP.
Then, to retrieve a token,
 the RP submits an authorization code
    and the PKCE verifier over anonymous networks \cite{tor,ohttp-rfc}, both of which are verified by the IdP,
     after anonymously authenticated to the IdP (i.e., along with a ring signature or privacy-pass token).
%On the contrary, the designs of RP anonymous authentication in ArpSSO \cite{ARPSSO} do not specially consider this full-exposure threat, even when we assume \emph{inexhaustible} RP domains and forgive the leakage of RP domain described in Appendix \ref{brute-force-domain}.
Then, the target RP anonymously retrieves the token,
        and any other RP cannot obtain the token from the IdP even if it intercepts the authorization code.
Either ring signature or privacy-pass token is much more efficient than zero-knowledge proofs \cite{ELPASSO} adopted in ArpSSO.

Finally, alternative to RP certificates \cite{uppresso-arxiv-all} and PS signatures \cite{PS-sig} sent from RPs,
        hashing-to-elliptic-curves \cite{irtf-cfrg-hash-to-curve-16} binds $ID_{RP}$ and the visited RP
            but with no cost of messages transmitted \cite{uppresso-arxiv-all}.
The design of RP certificates can also be utilized without extra messages transmitted:
    $ID_{RP}$ is enclosed in an RP's HTTPS/TLS certificate, signed by a trusted certification authority (CA) and used in secure communications between the RP and users.
Then, the web scripts directly obtain this certificate and $ID_{RP}$ \cite{Web-Certificate-API}.
This does not introduce extra trusted entities,
        because HTTPS/TLS is adopted in SSO services and the CA has been implicitly trusted by all entities.

%Finally,
%    in the designs of UppreSSO \cite{uppresso-conference,uppresso-arxiv}, the intentional decoupling of (\emph{a}) the identity transformations,
% (\emph{b}) the method to obtain the target RP's identity (e.g., RP certificate or hashing-to-elliptic-curves), (\emph{c}) the forwarding of identity tokens (e.g., the implicit flow or the authorization code flow),
% and (\emph{d}) the RP anonymous authentication to the IdP (e.g., implicit authentication, ring signature or privacy-pass token),
%    will facilitate the flexible integration of the identity-transformation approach in the future. 
\begin{table*}[htb]
    \caption{Functions and Parameters of Typical OPRFs}
    \centering
    \begin{tabular}{|c|c|c|c|c|} \hline
    & \textbf{HashDH} & \textbf{NR$_{\textrm{HE}}$} & \textbf{DY$_{\textrm{HE}}$} & \textbf{2HashRSA}\\ \hline

    \makecell{$z=\mathcal{PR}(k,x)$} & \makecell{$k \overset{{\scriptscriptstyle\$}}{\leftarrow} \mathbbm{F}_q$\\$x \overset{{\scriptscriptstyle\$}}{\leftarrow} \mathbbm{F}_q$\\$z=x^k$} & \makecell{$k=(a_0, a_1, \cdots, a_l) \overset{{\scriptscriptstyle\$}}{\leftarrow} \mathbbm{F}_q^{l+1}$\\$x=\tilde{x}_1\cdots \tilde{x}_l \in \{0,1\}^l$\\$z=g^{a_0\prod_{i=1}^{l}a_i^{\tilde{x}_i}}$} & \makecell{$k \overset{{\scriptscriptstyle\$}}{\leftarrow} \mathbbm{F}_q$\\$x \overset{{\scriptscriptstyle\$}}{\leftarrow} \mathbbm{F}_q$\\$z=g^{1/(k+x)}$} & \makecell{$(N,e,k) {\Leftarrow} RSA()$\\$x \overset{{\scriptscriptstyle\$}}{\leftarrow} \mathbbm{Z}_N$\\$z=H_2(x,x^k)$}\\ \hline

    $\omega$ & - & \makecell{$(r_1, \cdots, r_n) \overset{{\scriptscriptstyle\$}}{\leftarrow} \mathbbm{F}_q^l$\\$\omega=g^{a_0\prod_{i=1}^{l}1/r_i}$} & \makecell{$(sk,pk){\Leftarrow}AHE()$\\$\omega=Enc(k)$} & $\omega = e$ \\ \hline

    \makecell{$x'= \mathcal{BL}(x,t,\omega)$} & \makecell{$t \overset{{\scriptscriptstyle\$}}{\leftarrow} \mathbbm{F}_q$\\$x'=x^t$} & \makecell{$(sk,pk){\Leftarrow}AHE()$\\$x'= \{Enc(m_{i=1,\cdots, l})\}, m_i=(1-\tilde{x}_i,\tilde{x}_i)$} & \makecell{$t \overset{{\scriptscriptstyle\$}}{\leftarrow} \mathbbm{F}_q$\\$x'=(\omega Enc(x))^t$} & \makecell{$t \overset{{\scriptscriptstyle\$}}{\leftarrow} \mathbbm{Z}_N$\\$x'=xt^{\omega}$}\\ \hline

    \makecell{$z'= \mathcal{OPR}(k,x',\omega)$} & $z'=x'^k$ & $z'=\{Enc(m_i)^{(r_i,r_i a_i)}\}$ & \makecell{$t(k+x)=Dec(x')$\\$z'=g^{1/t(k+x)}$} & $z'=x'^{k}$ \\ \hline

    \makecell{$z= \mathcal{UBL}(z', x, t, \omega)$} & $z=z'^{1/t}$ & \makecell{$\{r_ia_i^{\tilde{x}_i}\}=Dec(z')$\\$z=\omega^{\prod_{i=1}^{l}r_ia_i^{\tilde{x}_i}}$} & $z=z'^t$ & $z=H_2(x,z'/t)$ \\ \hline

\end{tabular}
    \label{oprf-table}
{
\begin{enumerate}
\item   In 2HashRSA the calculations are conducted in $\mathbbm{Z}_N$, while those in other OPRFs are done in $\mathbbm{F}_q$.
\item  In the original HashDH, $z=H_1(x)^k$ and $H_1()$ is a collision-free hash function outputting uniformly random elements in $\mathbbm{F}_q$. As $x$ is randomly selected in $\mathbbm{F}_q$, we set $H_1(x)=x$.
Besides, it is easy to shift HashDH to elliptic curves, resulting in HashECDH, and the conclusions in this paper are applicable to HashECDH.
\item  $(sk, pk){\Leftarrow}AHE()$ generates a key pair of an additively HE scheme. % such as Paillier \cite{Paillier}.
 $Enc()$ and $Dec()$ represents its encryption and decryption, respectively. $Enc(a)Enc(b) = Enc(a+b)$ and $Enc(a_1,a_2)^{(b_1,b_2)} = Enc(a_1)^{b_1}Enc(a_2)^{b_2} = Enc(a_1b_1+a_2b_2)$.

\item   $(N,e,k) {\Leftarrow} RSA()$ generates an RSA key pair, and $H_2()$ is a collision-free hash function outputting uniformly random elements in $\mathbbm{Z}_N$.

\end{enumerate}}
\end{table*}

\section{Constructing Identity Transformations Based on OPRFs}
\label{idt-oprf}
In this section, we uncover the relationship between the identity transformations in generalized UppreSSO and general OPRFs.

\subsection{Basic Properties of OPRFs}
\label{OPRF-Formal}
Given a pseudo-random function $z=\mathcal{PR}(k,x)$, where $k$ is the \emph{secret key} held by an OPRF server
    and $x$ is a \emph{private input} from an OPRF user,
    the OPRF server and the user cooperate to calculate $z$ as below \cite{sok-oprf,strong-oprf,eff-psi-oprf,voprf-proved,oprf-ot-si,two-sided-oprf}:
\begin{enumerate}
\item
An \emph{optional} parameter $\omega$ is selected by the IdP.
\item
An OPRF user blinds her input $x$ into $x' = \mathcal{BL}(x, t, \omega)$ using a random number $t$, and sends $x'$ to the OPRF server.
\item
The OPRF server calculates the output as $z'= \mathcal{OPR}(k, x',\omega)$.
\item
On receiving $z'$, the user unblinds it to $z = \mathcal{UBL}(z',x,t,\omega)$.
\end{enumerate}

As shown in Table \ref{oprf-table}, the formalization covers typical protocols: (\emph{a}) HashDH \cite{strong-oprf,eff-psi-oprf}, 
    (\emph{b}) the NR OPRF using homomorphic encryption (HE) \cite{robust-pp-ss,voprf-proved}, denoted as NR$_{\textrm{HE}}$ in this paper, (\emph{c}) the DY OPRF using HE \cite{oprf-ot-si,oprf-deduplication,two-sided-oprf}, denoted as DY$_{\textrm{HE}}$,
    and (\emph{d}) 2HashRSA \cite{voprf-proved}.
Meanwhile, $\omega$ plays different roles in these protocols:
 (\emph{a}) $\omega$ does not exist in HashDH;
(\emph{b}) in NR$_{\textrm{HE}}$ $\omega=g^{a_0\prod_{i=1}^{l}1/r_i}$ is a parameter of both $\mathcal{OPR()}$ and $\mathcal{UBL}()$,
        and (\emph{c}) in DY$^{\textrm{HE}}$ and 2HashRSA it is processed only in $\mathcal{BL}()$.

An OPRF protocol generally satisfies the basic properties \cite{sok-oprf,strong-oprf}:
\begin{itemize}
\item \textbf{Pseudo-Randomness:}
In an OPRF user's view, $z$ is indistinguishable from uniformly random variables,
    and she learns nothing on $k$.
\item \textbf{Correctness:} For any $x$, $k$, $t$, and $\omega$, $\mathcal{UBL}(z',x,t,\omega)$ is equal to $\mathcal{PR}(k,x)$.
\item \textbf{Obliviousness:}
The OPRF server learns nothing on $x$.
\end{itemize}

\subsection{Building Privacy-Preserving SSO Services Based on OPRFs}
\label{relationship-idt-oprf}
The ($\mathbbm{F}_q$ versions of the) identity transformations in UppreSSO \cite{uppresso-arxiv-all}
\emph{mathematically} utilize the same functions as the OPRF of HashDH \cite{voprf-proved,oprf-proved};
    that is,
four functions of HashDH (i.e., $\mathcal{PR}()$, $\mathcal{BL}()$, $\mathcal{OPR}()$, and $\mathcal{UBL}()$), work as
 (the $\mathbbm{F}_q$ versions of) 
 $\mathcal{F}_{Acct\ast}()$, $\mathcal{F}_{PID_{RP}}()$, $\mathcal{F}_{PID_U}()$, and $\mathcal{F}_{Acct}()$ in UppreSSO \cite{uppresso-arxiv-all}, respectively.

Next, we try to construct identity transformations 
for privacy-preserving SSO, based on a general OPRF scheme other than HashDH.
Given an OPRF protocol conforming to the formalization in Section \ref{OPRF-Formal},
   a generalized UppreSSO system is built as below, which adopts the same system model as UppreSSO does in Section \ref{system-model}.

Unique $ID_U = k$ and $ID_{RP} = x$ are randomly selected and assigned to a registered user and a registered  RP by the honest-but-curious IdP, respectively.
 $Acct=\mathcal{F}_{Acct\ast}(ID_{U}, ID_{RP}) = \mathcal{PR}(k,x) =z$ is then automatically assigned to a registered user at each RP.
The IdP, RPs, and users follow these specifications in a login.
\begin{enumerate}
  \item A user visits an RP identified as $ID_{RP}=x$, 
    and she receives $\omega$ from the IdP if needed.
  \item The user requests an identity  token for $PID_{RP}$ from the IdP,
    after randomly selecting $t$ and calculating $PID_{RP} = \mathcal{F}_{PID_{RP}}(ID_{RP}, t, \omega) = \mathcal{BL}(x, t, \omega) = x'$.
  \item After authenticating the user as $k$,
         the IdP calculates $PID_{U} = \mathcal{F}_{PID_U}(ID_U, PID_{RP},\omega) = \mathcal{OPR}(k, x',\omega) = z'$, and signs $TK$ binding $PID_{RP}$ and $PID_{U}$ (i.e., $x'$ and $z'$).
    The IdP also signs $\omega$ in $TK$, if it is needed by the RP to derive accounts
    (i.e., $\omega$ is a parameter of $\mathcal{UBL}()$, or it is processed in $\mathcal{BL}()$ and $PID_{RP}$ is checked by the RP).

  \item $TK$ is forwarded by the user as well as $t$ to the RP in the implicit flow,
        or in the authorization code flow it is retrieved by the RP 
        after anonymously authenticated to the IdP.
  \item
  If $TK$ is well signed and unexpired, the RP derives $Acct = \mathcal{F}_{Acct}(PID_{U}, ID_{RP}, t, \omega) = \mathcal{UBL}(z', x,t, \omega) = z$.
The token holder is allowed to log in as $Acct$, if it is meaningful.
\label{pid-rp-check-step}
\end{enumerate}

In Step \ref{pid-rp-check-step},
    according to the properties of the underlying OPRF (see Section \ref{security-properties-list} for more details),
     the RP may accept (\emph{a}) \emph{any} signed tokens or (\emph{b}) \emph{only} the ones enclosing $PID_{RP}$ equal to $[t]ID_{RP}$.

Table \ref{tbl:oprf2idtsso} compares the functions.
Compared with the original design of UppreSSO \cite{uppresso-arxiv-all},
    the parameters of  $\mathcal{F}_{Acct\ast}()$, $\mathcal{F}_{PID_{RP}}()$, $\mathcal{F}_{ID_U}()$, and $\mathcal{F}_{Acct}()$ are slightly revised according to those of $\mathcal{PR}()$, $\mathcal{BL}()$, $\mathcal{OPR}()$, and $\mathcal{UBL}()$ in typical OPRFs.
Sometimes $\omega$ depends on $k$, and then is determined after a user is authenticated, e.g., $\omega = Enc(k)$ in DY$_{\textrm{HE}}$ and $\omega = e$ in 2HashRSA.
In this case, the IdP first authenticates the user as $k$, and sends $\omega$ to the user.

\begin{table}[bt]
    \caption{Identity Transformations in SSO vs. OPRFs}
    \centering{
    \begin{tabular}{|r|l|} \hline
    \textbf{Identity Transformations} & \textbf{OPRF} \\ \hline
%    \multicolumn{2}{|c|}{\textbf{Variable}} \\ \hline
%    $ID_U$ & $k$ \\ \hline
%    $ID_{RP}$ & $x$ \\ \hline
%    \multicolumn{2}{|c|}{\textbf{Function}} \\ \hline
    $Acct = \mathcal{F}_{Acct\ast}(ID_U, ID_{RP})$ & $z = \mathcal{PR}(k,x)$ \\ \hline
    $PID_{RP} = \mathcal{F}_{PID_{RP}}(ID_{RP}, t, \omega)$ & $x' = \mathcal{BL}(x,t,\omega)$ \\ \hline
    $PID_{U} = \mathcal{F}_{PID_{U}}(ID_{U}, PID_{RP},\omega)$ & $z' = \mathcal{OPR}(k,x',\omega)$ \\ \hline
    $Acct = \mathcal{F}_{Acct}(PID_U, ID_{RP}, t, \omega)$ & $z = \mathcal{UBL}(z',x,t,\omega)$ \\ \hline
\end{tabular}}
    \label{tbl:oprf2idtsso}
\end{table}

Let $\mathbbm{ID}_{RP}  = \{ID_{RP_{j=1,\cdots,p}}\}$, $\mathbbm{ID}_U = \{ID_{U_{i=1,\cdots,s}}\}$,
    $\mathbbm{x} = \{x_1, \cdots, x_p\}$ and $\mathbbm{k} = \{k_1, \cdots, k_s\}$.
Then, $\mathbbm{A}cct  = \{Acct_{i,j=1,\cdots,p}|_{i=1,\cdots,s}\} = \{\mathcal{F}_{Acct\ast}(ID_{U_i}, ID_{RP_j})\}$ and $\mathbbm{z} = \{z_{i,j=1,\cdots,p}|_{i=1,\cdots,s}\}=\{\mathcal{PR}(k_i,x_j)\}$.
Therefore, a generalized UppreSSO system running with $\mathbbm{ID}_{RP}$, $\mathbbm{ID}_U$ and $\mathbbm{A}cct$, corresponds to an OPRF system consisting of
    an OPRF server and $p$ OPRF users.
Each OPRF user has a private input $x_j$,
the OPRF server holds $s$ keys, and they cooperate to calculate $z$:
    A user blinds its input into $x'=\mathcal{BL}(x_j, t, \omega)$,
        the OPRF server \emph{randomly} picks up $k_i \in \mathbbm{k}$ and returns $z'=\mathcal{OPR}(k_i, x',\omega)$, and the user finally receives $z_{i,j}=\mathcal{UBL}(z', x,t, \omega)$.

\subsection{Properties Required in Generalized UppreSSO Systems}
\label{property-oprf}

We analyze the required properties of identity transformations in a generalized UppreSSO (or the underlying OPRF) as below.
%Our analysis significantly extends the understanding of security and privacy in UppreSSO \cite{uppresso-arxiv,uppresso-conference,ARPSSO}.

\subsubsection{Threat Model and Adversarial Scenarios}
$ID_{RP}=x$ and $Acct=z$ are publicly known,
    but $k$ cannot be deduced because $z=\mathcal{PR}(k,x)$ is a pseudo-random function.
$ID_U$ is known to only the IdP;
 otherwise, RP unlinkability is broken: Colluding RPs 
    could calculate $Acct=\mathcal{F}_{Acct\ast}(ID_{U}, ID_{RP})=\mathcal{PR}(k,x)$
        for each known $ID_U$ at these RPs and accordingly link the accounts.
Fortunately, $ID_U=k$ is processed only by the IdP internally, not enclosed in any message or leaked in the underlying OPRF protocol.

To ensure the security guarantees of authentication services provided by \emph{an honest IdP},
    user identification and RP designation are proved against \emph{malicious RPs colluding with some users} \cite{uppresso-arxiv-all},
                which attempt to (\emph{a}) impersonate an honest user to log into some honest RP
   or (\emph{b}) entice an honest user to log into an honest RP as another's account,
by manipulating $t$ and/or sending $TK$ to an RP not designated.

We explain the corresponding attacks in the underlying OPRF system with \emph{an honest OPRF server}:
By manipulating $t$ and/or sending $(x', z')$ to another \emph{honest} OPRF user,
    a \emph{malicious} OPRF user attempts to (\emph{a}) calculate \emph{another} output but \emph{meaningful} for itself
        or (\emph{b}) entice the victim \emph{honest} user to accept some \emph{meaningful} output.

We analyze the privacy properties as follows \cite{uppresso-arxiv-all}.
\emph{An honest-but-curious IdP} is considered in IdP untraceability,
 while RP unlinkability is analyzed against \emph{a group of semi-honest RPs and users},
 which do not actively violate the protocol
    but share their information, attempting to link an honest user's login activities.
No malicious RP or user is considered in the proofs of privacy \cite{uppresso-arxiv-all}.\footnote{If the initiating user or the RP deviates from the specifications, no meaningful account is derived
and it makes no sense to track a login resulting in meaningless accounts.}

Thus, two privacy guarantees of the underlying OPRF are analyzed
    against \emph{an honest-but-curious OPRF server} and \emph{a group of semi-honest OPRF users}, respectively.
That is, the \emph{honest-but-curious} OPRF server attempts to learn about the input,
while the group of semi-honest users attempts to infer whether different keys or not are used in two protocol instances initiated by different users.\footnote{In generalized UppreSSO there are even some semi-honest users sharing information with the adversaries, and this means that in the underlying OPRF the group of semi-honest adversaries are agranted supplementary information on some OPRF keys but
    their attacks aim at other uncompromised keys. See Section \ref{privacy-properties-list} for more details.}

%We assume (\emph{a})
%authenticity, confidentiality and integrity of an identity token, and (\emph{b}) no direct privacy leakage in the requesting, receiving and forwarding of identity tokens.
%These assumptions are realized by the designs specific for web applications in UppreSSO \cite{uppresso-conference}, compatible with OIDC.

\subsubsection{Basic properties}
In order to provide non-anonymous authentication services as regular SSO protocols,
    in generalized UppreSSO it requires \emph{account uniqueness}:
    For any $ID_{RP} \in \mathbbm{ID}_{RP}$,
if $ID_{\hat{U}} \neq ID_{\check{U}}$,
      then $\mathcal{F}_{Acct\ast}(ID_{\hat{U}}, ID_{RP}) \neq \mathcal{F}_{Acct\ast}(ID_{\check{U}}, ID_{RP})$ where $ID_{\hat{U}},ID_{\check{U}} \in \mathbbm{ID}_U$.
Thus, in the underlying OPRF, it requires that,
    for any $x \in \mathbbm{x}$,
      $\mathcal{PR}(\hat{k}, x) \neq \mathcal{PR}(\check{k}, x)$
      if $\hat{k} \neq \check{k}$ where $\hat{k},\check{k} \in \mathbbm{k}$, which is called \emph{output uniqueness} of OPRFs in this paper.

\emph{Account correctness} of generalized UppreSSO in the case of no attack,
    i.e., $\mathcal{F}_{Acct}(PID_U, ID_{RP}, t, \omega) = \mathcal{F}_{Acct\ast}(ID_{U}, ID_{RP})$, is ensured by
\emph{correctness} of the underlying OPRFs, i.e., $\mathcal{UBL}(z',x,t,\omega) = \mathcal{PR}(k,x)$.

\subsubsection{Security properties}
\label{security-properties-list}
In the case of malicious RPs and users,
to implement secure authentication services,
\emph{an identity token requested by a user to visit an honest RP,
    enables only this user to log into only this RP as her account at it}.
Note that it makes no sense to discuss the login results at a malicious RP.
%This produces \emph{user identification} and \emph{RP designation}, as briefly mentioned in Section \ref{uppresso-security}.

An RP may derive accounts based on (\emph{a}) \emph{any} signed identity tokens, 
    as demonstrated in UppreSSO \cite{uppresso-arxiv-all},
    or (\emph{b}) 
\emph{only} tokens binding \emph{matching} $PID_{RP}$, as implied by the original OIDC protocol \cite{OpenIDConnect} and other privacy-preserving SSO solutions \cite{BrowserID,SPRESSO,NIST2017draft,POIDC,save-flow,up-sso,miso}
    where only tokens binding matching RP (pseudo-)identities are accepted.
These different RP operations result in \emph{alternative} variations of requirements:
    Security properties \emph{w/o $PID_{RP}$ checking} and those \emph{w/ $PID_{RP}$ checking}.

Generalized UppreSSO systems can be built based on an OPRF with any variation of these security properties.
 %   provided that the corresponding RP operations are specified in Step \ref{pid-rp-check-step} of the SSO protocol.
If user identification and RP designation w/o $PID_{RP}$ checking are ensured,
 an RP accepts \emph{any} signed tokens to derive meaningful accounts;
    alternatively, if these properties are ensured only with $PID_{RP}$ checking, an RP accepts \emph{only} signed tokens binding \emph{matching} $PID_{RP}$.
%These variations significantly extend the identity-transformation approach of UppreSSO.
%The more efficient protocol operations are recommended,
% if the stronger properties (i.e., user identification and RP designation w/o $PID_{RP}$ checking) are ensured.

Based on the explanations of user identification and RP designation in Section \ref{uppresso-security},
    the security properties w/o $PID_{RP}$ checking are formally defined as below.

\noindent\textbf{User Identification w/o $PID_{RP}$ Checking:} 
Given known $\mathbbm{A}cct$ and unknown $\mathbbm{ID}_U$,
for any $ID_{RP} \in \mathbbm{ID}_{RP}$, 
     adversaries cannot find $\check{t}$ and $TK$ binding $PID_{RP} = \mathcal{F}_{PID_{RP}}(ID_{RP}, \hat{t}, \omega)$
     and $PID_{\hat{U}}=\mathcal{F}_{PID_U}(ID_{\hat{U}}, PID_{RP}, \omega)$ which satisfy that $\mathcal{F}_{Acct}(PID_{\hat{U}}, ID_{RP}, \check{t}, \omega) = \mathcal{F}_{Acct\ast}(ID_{\check{U}}, ID_{RP})$, where $ID_{\hat{U}} \neq ID_{\check{U}}$ and $ID_{\hat{U}},ID_{\check{U}} \in \mathbbm{ID}_U$.

This property means that,
        when $TK$ is requested by some user $\hat{U}$ for $PID_{RP}$ which is calculated based on $ID_{RP}$,
    adversaries cannot exploit $TK$ to impersonate another user $\check{U}$ at this designated RP by manipulating $\check{t}$.
That is, based on $TK$ the designated honest RP will derive either (\emph{a}) the account belonging to the user requesting $TK$ if $\check{t} = \hat{t}$ 
or (\emph{b}) meaningless accounts if $\check{t} \neq \hat{t}$.

The corresponding property is proposed for the underlying OPRF,
called \emph{strong collision resistance on 1st-input}:
Given known $\mathbbm{z}$ and unknown $\mathbbm{k}$,
    for any $x \in \mathbbm{x}$,
adversaries cannot find $\check{t}$ and some protocol instance $(x, \hat{x}', \hat{z}', \hat{t}, \omega)$ generated with unknown $\hat{k}$
which satisfy that $\mathcal{UBL}(\hat{z}', x,\check{t}, \omega) = \mathcal{PR}(\check{k}, x)$, where $\hat{k} \neq \check{k}$ and $\hat{k}, \check{k} \in \mathbbm{k}$.

\noindent\textbf{RP Designation w/o $PID_{RP}$ Checking:}
Given known $\mathbbm{ID}_{RP}$, known $\mathbbm{A}cct$ and unknown $\mathbbm{ID}_U$,
 it is computationally impossible for adversaries to find ${j_1}$, ${j_2}$, ${i_1}$, ${i_2}$, $t_1$, $t_2$, and $\omega$ which satisfy that $\mathcal{F}_{Acct}(\mathcal{F}_{PID_U}(ID_{U_{i_1}}, PID_{RP}, \omega), ID_{RP_{j_2}}, t_2, \omega) = \mathcal{F}_{Acct\ast}(ID_{U_{i_2}}, ID_{RP_{j_2}})$,
where ${j_1} \neq {j_2}$, $1 \leq j_1, j_2 \leq p$, $1 \leq i_1, i_2 \leq s$,
 and $PID_{RP}$ may be calculated as $\mathcal{F}_{PID_{RP}}(ID_{RP_{j_1}}, t_1, \omega)$
    or arbitrarily generated (i.e., the $j_1$-th RP is designated or no RP).

This means that,
    adversaries cannot request a token for $PID_{RP}$ which is calculated based on $ID_{RP_{j_1}}$ or even arbitrarily generated,
        but exploit this token to log into the $j_2$-th RP (${j_1} \neq {j_2}$) as a meaningful account.
When $PID_{RP}$ is arbitrarily generated by adversaries,
    the IdP signs a token as usual but no registered RP is designated.
That is, 
    at any honest RPs other than the designated one, no meaningful account will be derived based on $TK$.

\emph{Strong collision resistance on 2nd-input} is proposed:
Given known $\mathbbm{x}$, known $\mathbbm{z}$ and unknown $\mathbbm{k}$, 
     adversaries cannot find ${j_1}$, ${j_2}$, ${i_1}$, ${i_2}$, $t_1$, $t_2$, and $\omega$ satisfying that 
        $\mathcal{UBL}(\mathcal{OPR}(k_{i_1},x',\omega), x_{j_2}, t_2, \omega) = \mathcal{PR}(k_{i_2}, x_{j_2})$,
where ${j_1} \neq {j_2}$, $1 \leq j_1, j_2 \leq p$, $1 \leq i_1, i_2 \leq s$,
    and $x'$ is calculated as $\mathcal{BL}(x_{j_1}, t_1, \omega)$ or arbitrarily generated.

On the contrary, when an (honest) RP accepts only tokens binding \emph{matching} $PID_{RP}$, %i.e., $PID_{RP} = \mathcal{F}_{PID_{RP}}(ID_{RP}, t, \omega)$ where $ID_{RP}$ is its identity.
the security properties are revised as below.

\noindent\textbf{User Identification w/ $PID_{RP}$ Checking:}
Given known $\mathbbm{A}cct$ and unknown $\mathbbm{ID}_U$,
    for any $ID_{RP} \in \mathbbm{ID}_{RP}$,
     adversaries cannot find $\check{t}$ and $TK$ binding $PID_{RP} = \mathcal{F}_{PID_{RP}}(ID_{RP}, \hat{t}, \omega)$ and $PID_{\hat{U}}=\mathcal{F}_{PID_U}(ID_{\hat{U}}, PID_{RP}, \omega)$
     satisfying that $\mathcal{F}_{PID_{RP}}(ID_{RP}, \check{t}, \omega) = PID_{RP}$
     and $\mathcal{F}_{Acct}(PID_{\hat{U}}, ID_{RP}, \check{t}, \omega) = \mathcal{F}_{Acct\ast}(ID_{\check{U}}, ID_{RP})$, where $ID_{\hat{U}} \neq ID_{\check{U}}$ and $ID_{\hat{U}},ID_{\check{U}} \in \mathbbm{ID}_U$.

As an RP additionally checks $PID_{RP}$ enclosed in a received token,
    in order to impersonate $\check{U}$,
    adversaries need to manipulate $\check{t}$ 
        which deceive the RP to accept $TK$ (i.e., $\mathcal{F}_{PID_{RP}}(ID_{RP}, \check{t}, \omega)$ is equal to $PID_{RP}$) and derive the account belonging to $\check{U}$
         (i.e., $\mathcal{F}_{Acct}(PID_{\hat{U}}, ID_{RP}, \check{t}, \omega)$ derives a meaningful account).

The property of \emph{weak collision resistance on 1st-input} for OPRFs is defined:
Given known $\mathbbm{z}$ and unknown $\mathbbm{k}$,
    for any $x \in \mathbbm{x}$,
     adversaries cannot find $\check{t}$ and some protocol instance $(x, \hat{x}', \hat{z}', \hat{t}, \omega)$ generated with unknown $\hat{k}$
     which satisfy that 
        $\mathcal{BL}(x, \check{t}, \omega) = \hat{x}'$
        and $\mathcal{UBL}(\hat{z}', x, \check{t}, \omega) = \mathcal{PR}(\check{k}, x)$,
      where $\check{k} \neq \hat{k}$ and $\check{k}, \hat{k} \in \mathbbm{k}$.

\noindent\textbf{RP Designation w/ $PID_{RP}$ Checking:}
Given known $\mathbbm{ID}_{RP}$, known $\mathbbm{A}cct$ and unknown $\mathbbm{ID}_U$,
 adversaries cannot find ${j_1}$, ${j_2}$, ${i_1}$, ${i_2}$, $t_1$, $t_2$, and $\omega$ which satisfy that 
$\mathcal{F}_{PID_{RP}}(ID_{RP_{j_2}}, t_2, \omega) = PID_{RP}$
and $\mathcal{F}_{Acct}(\mathcal{F}_{PID_U}(ID_{U_{i_1}}, PID_{RP}, \omega), ID_{RP_{j_2}}, t_2, \omega) = \mathcal{F}_{Acct\ast}(ID_{U_{i_2}}, ID_{RP_{j_2}})$, 
where ${j_1} \neq {j_2}$, $1 \leq j_1, j_2 \leq p$, $1 \leq i_1, i_2 \leq s$,
        and $PID_{RP}$ may be calculated as $\mathcal{F}_{PID_{RP}}(ID_{RP_{j_1}}, t_1, \omega)$
    or arbitrarily generated.

\emph{Weak collision resistance on 2nd-input} is proposed for OPRFs:
Given known $\mathbbm{x}$, known $\mathbbm{z}$ and unknown $\mathbbm{k}$,
adversaries cannot find ${j_1}$, ${j_2}$, $i_1$, $i_2$, $t_1$, $t_2$, and $\omega$ satisfying that $\mathcal{BL}(x_{j_2}, t_2, \omega) = x'$ and $\mathcal{UBL}(\mathcal{OPR}(k_{i_1}, x', \omega), x_{j_2}, t_2, \omega) = \mathcal{PR}(k_{i_2}, x_{j_2})$,
 where ${j_1} \neq {j_2}$, $1 \leq j_1, j_2 \leq p$, $1 \leq i_1, i_2 \leq s$,
        and $x'$ is calculated as $\mathcal{BL}(x_{j_1}, t_1, \omega)$ or arbitrarily generated.

Finally, when analyzing the properties as above,
we do not particularly consider the manipulation of $\omega$ by adversaries.
\begin{itemize}
  \item In HashDH $\omega$ does not exist.
  \item In generalized UppreSSO built based on NR$_{\textrm{HE}}$, $\omega$ is randomly generated by the IdP and signed in $TK$, as described in Section \ref{relationship-idt-oprf}. So adversaries cannot manipulate it.
  \item In a system based on DY$_{\textrm{HE}}$ or 2HashRSA,
    $\omega$ is not signed in $TK$ but it is another form of $k$ (i.e., $\omega = Enc(k)$ in DY$_{\textrm{HE}}$ and $\omega = e$ in 2HashRSA), so the impacts of $\omega$ have been considered as we consider $\hat{k} \neq \check{k}$ (or $k_{i_1} \neq k_{i_2}$) in the proofs of collision resistance.
\end{itemize}

\subsubsection{Privacy properties}
\label{privacy-properties-list}
Both IdP-based login tracing and RP-based identity linkage
    are prevented in generalized UppreSSO.

\noindent\textbf{IdP Untraceability:}
\emph{The honest-but-curious IdP learns nothing on $ID_{RP}$ from $PID_{RP}$.}

In a generalized UppreSSO system,
        among the messages received by the honest-but-curious IdP,
only $PID_{RP}$ is calculated based on $ID_{RP}$.
% So IdP untraceability is defined as below.
This definition means that, in the underlying OPRF,
 the honest-but-curious server learns nothing on $x$ from $x'$.
Thus, IdP untraceability is equivalent to \emph{obliviousness} of OPRFs.

\noindent \textbf{RP Unlinkability:}
\emph{The logins visiting different semi-honest RPs by any honest users, are indistinguishable to these RPs, even when they collude with other users.}

If the logins visiting different RPs by any honest users are indistinguishable to these semi-honest RPs,
 they cannot link the logins by a certain user across them.

Each login corresponds to an r-tuple of the messages received from the OPRF server in a protocol instance of the underlying OPRF.
 %    which is finished by some user and the OPRF server randomly picking up a secret key.
In generalized UppreSSO built based on HashDH,
    each login is expressed as $(x, x', z', t)$;
or, when it is built based on NR$_{\textrm{HE}}$, DY$_{\textrm{HE}}$ and 2HashRSA,
    a login is expressed as $(x, x', z', t, \omega)$,
    $(x, x', z', t)$ and $(x, x', z', t)$, respectively, because $\omega$ is a parameter of $\mathcal{UBL}()$ only in NR$_{\textrm{HE}}$.
Besides, if $PID_{RP}$ is checked by an RP before accepting a token,
    the parameters of $\mathcal{BL}()$ are also included in each r-tuple.
% Besides, the information shared among colluding RPs and users of a generalized UppreSSO system can be mapped to the tuples collected by this OPRF user.

While it has been proved from scratch in UppreSSO built based on Hash(EC)DH \cite{uppresso-arxiv-all,ARPSSO},
RP unlinkability is equivalent to \emph{indistinguishability of keys} of the underlying OPRF:
    After initiating the OPRF protocol with different inputs,
based on the r-tuples the group of semi-honest OPRF users cannot tell whether the OPRF server uses different keys or not across these protocol instances,
    even when they are granted some \emph{supplementary} sets of r-tuples.
All r-tuples in a supplementary set are generated with one unknown OPRF key,
and each supplementary set of r-tuples 
    can be mapped to an SSO user colluding with RPs in generalized UppreSSO.

For an OPRF scheme, indistinguishability of keys implies pseudo-randomness;
    otherwise,
        OPRF users could learn something on the key after a protocol instance,
            which could be exploited to distinguish different keys (i.e., break RP unlinkability).

On the other hand,
    pseudo-randomness does not ensure indistinguishability of keys.
For example, 
in 2HashRSA indistinguishability of keys is not ensured,
because $N$ usually uniquely identifies an OPRF key
    but it is a public parameter of $\mathcal{UBL}(z',x,t,\omega)=H_2(x,z'/t \bmod N)$;
or, when the server generates one HE key pair for multiple OPRF keys in DY$_{\textrm{HE}}$,
        $\omega = Enc(k)$ identifies a key unless a probabilistic HE scheme is adopted.

\emph{Strong key-identifier freeness} of OPRFs is proposed in this paper:
All messages (i.e., $z'$ and $\omega$) and the public parameters (e.g., $q$ and $N$)
cannot be exploited to distinguish an OPRF key from others.
Meanwhile, \emph{weak key-identifier freeness} is also proposed:
The parameters of $\mathcal{UBL}()$ and the public parameters cannot be exploited to distinguish a key.

%Indistinguishability of multiple OPRF keys is equivalent to both pseudo-randomness and key-identifier freeness of an OPRF.
%If strong key-identifier freeness is ensured,
%    RP-based identity linkage is prevented whether $PID_{RP}$ is checked by an RP or not in Step \ref{pid-rp-check-step}.
%If weak (but not strong) key-identifier freeness is ensured,
%    RP unlinkability is broken for some arguments of $\mathcal{BL}()$ identifies an OPRF key
%     unless $PID_{RP}$ is not checked by an RP before accepting a token.

These extended properties of OPRFs,  
namely \emph{key-identifier freeness} and \emph{collision resistance on 1st/2nd-input},
   have not be investigated in the literature \cite{sok-oprf}.
%Not all OPRFs are qualified for a generalized UppreSSO system, because these properties are not always ensured.

\subsection{Proofs of SSO-Related Properties}
\label{proofs-oprf}

First of all, in the typical OPRFs \emph{account correctness} and \emph{IdP untraceability}
    are ensured,
        because they are equivalent to \emph{correctness} and \emph{obliviousness} of OPRFs, respectively.

\emph{RP unlinkability} is equivalent to \emph{indistinguishability of keys} (i.e., both pseudo-randomness and key-identifier freeness) of an OPRF.
This property is analyzed as below and some OPRF protocols are revised slightly.
\begin{itemize}
  \item Multiple keys are indistinguishable in HashDH and NR$_{\textrm{HE}}$ where pseudo-randomness is ensured \cite{strong-oprf,voprf-proved},
   if there is an \emph{identical} finite field $\mathbbm{F}_q$ for all OPRF keys.
  \item It is ensured in DY$_{\textrm{HE}}$,
        if (\emph{a}) the server adopts an \emph{identical} finite field $\mathbbm{F}_q$
         and (\emph{b}) $\omega=Enc(k)$ is \emph{refreshed} in each protocol instance.
This require that the additively HE key pair is \emph{ephemeral}
        or the adopted HE scheme is \emph{probabilistic}.\footnote{If a \emph{deterministic} additively HE scheme is adopted and the HE key pair is \emph{permanent}, $\omega=Enc(k)$ is restored by calculating $x'^{1/t}/Enc(x)$ and it uniquely identifies an OPRF key.} 
% Anyway, the widely-used additively HE scheme, Paillier \cite{Paillier}, is probabilistic.}
  \item In 2HashRSA pseudo-randomness is ensured \cite{voprf-proved} but not
  key-identifier freeness, because $N$ uniquely identifies a key.
We slightly revise it as follows, denoted as 2HashRSA$_N$ in this paper:
    An OPRF server generates \emph{multiple key pairs} but for an \emph{identical} $N$,
        i.e., $e_ik_i =1 \bmod \phi(N)$ for every OPRF key.
Then, weak but not strong key-identifier freeness is ensured in 2HashRSA$_N$,
    because $e$ is a parameter of $\mathcal{BL}()$ and uniquely identifies a key.
\end{itemize}

We prove the other SSO-related properties (i.e., \emph{output uniqueness} and \emph{collision resistance on 1st/2nd-input}) of these revised protocols.

\begin{theorem}
Output uniqueness is ensured in HashDH, DY$_{\textrm{HE}}$ and 2HashRSA$_N$,
    but not in NR$_{\textrm{HE}}$.
\end{theorem}

\noindent\textbf{Proof.}
This property is ensured in HashDH and DY$_{\textrm{HE}}$.
In HashDH, $k \overset{{\scriptscriptstyle\$}}{\leftarrow} \mathbbm{F}_q$, $x \overset{{\scriptscriptstyle\$}}{\leftarrow} \mathbbm{F}_q$,  and $z=x^k$. When $x$ is selected as a generator of $\mathbbm{F}_q$,
    $z$ is a bijective function of $k$.
In DY$_{\textrm{HE}}$,
    $k \overset{{\scriptscriptstyle\$}}{\leftarrow} \mathbbm{F}_q$, $x \overset{{\scriptscriptstyle\$}}{\leftarrow} \mathbbm{F}_q$, and $z=g^{1/(k+x)}$.
Given $x$, $1/(k+x)$ is a bijective function of $k$.
Because $g$ is a generator of $\mathbbm{F}_q$, $g^{1/(k+x)}$ is also a bijective function of $k$.

In 2HashRSA$_N$,
$(e,k) \overset{{\scriptscriptstyle\$}}{\Leftarrow} RSA(N)$, $z=H_2(x,x^k)$ and $|z|=N$.
For any $x \in \mathbbm{x}$, if $\hat{k} \neq \check{k}$, the probability that $H_2(x, x^{\hat{k}}) = H_2({x}, x^{\check{k}})$ is negligible, due to collision-freeness of $H_2()$ and security of RSA.

% where $k=(a_0, a_1, \cdots, a_n) \overset{{\scriptscriptstyle\$}}{\leftarrow} \mathbbm{F}_q^{n+1}$, $x=\tilde{x}_1\cdots \tilde{x}_n \in \{0,1\}^n$, and $z=g^{a_0\prod_{i=1}^{n}a_i^{\tilde{x}_i}}$,
 Because $|\mathbbm{k}| = q^{n+1}$ but $|\mathbbm{z}|=q$ in NR$_{\textrm{HE}}$,
sometimes $\hat{z}$ is equal to $\check{z}$ 
 when $\hat{k} \neq \check{k}$.
So output uniqueness is not ensured in NR$_{\textrm{HE}}$.
The NR OPRF using obviously transfer (OT) \cite{strong-oprf,eff-psi-oprf} does not strictly conform to the formalization in Section \ref{OPRF-Formal}, but this conclusion is also applicable to it.
\hfill $\square$
\vspace{2mm}

In the remainder we discuss HashDH, DY$_{\textrm{HE}}$, and 2HashRSA$_N$, but not NR$_{\textrm{HE}}$.
Moreover, in the proofs of Theorems \ref{theorem-uid-wo-checking} and \ref{theorem-rp-wo-checking} on 2HashRSA$_N$,
    we assume $x_j^{k_i}$ is kept \emph{secret} and known to only the $j$-th RP (and also the IdP).

%In the proofs of Theorems \ref{theorem-uid-wo-checking} and \ref{theorem-rp-wo-checking}, 
%        we assume that $x_j^{k_i}$ is kept secret to adversaries in 2HashRSA$_N$.
%In a generalized UppreSSO system built based on 2HashRSA$_N$, $z= H_2(x_j,z'/t) = H_2(x_j,x_j^{k_i})$ is calculated by an RP internally
%    and $x_j^{k_i}$ cannot be deduced based on public $z$ and $x_j$; that is, it is known to only this RP (and also the IdP).
%Meanwhile, Theorems \ref{theorem-uid-w-checking-rsa} and \ref{theorem-rp-w-checking-rsa} consider
%    publicly-known $x_j^{k_i}$ in 2HashRSA$_N$.

\begin{theorem}
Strong collision resistance on 1st-input is ensured in HashDH, DY$_{\textrm{HE}}$,
    and 2HashRSA$_N$ with secret $x_j^{k_i}$.
\label{theorem-uid-wo-checking}
\end{theorem}

\noindent\textbf{Proof.}
In HashDH it requires that, for any $x\in\{\mathbbm{x}\}$, adversaries cannot find $\hat{t}$ and $\check{t}$ satisfying that $x^{\hat{t}\hat{k}/\check{t}} = x^{\check{k}}$ where $\check{k} \neq \hat{k}$.
When $\mathbbm{z}=\{x_j^{k_i}\}$ is publicly known but $\mathbbm{k}=\{k_i\}$ is unknown, this is equivalent to the discrete logarithm problem (DLP) (i.e., to find $\hat{t}/\check{t} = \log_{x^{\hat{k}}}{x^{\check{k}}} = \log_{\hat{k}}{\check{z}}$).

In DY$_{\textrm{HE}}$ it requires that, for any $x$, 
     adversaries cannot find $\hat{t}$ and $\check{t}$ satisfying that $g^{\check{t}/\hat{t}(\hat{k}+x)} = g^{1/(\check{k}+x)}$ where $\check{k} \neq \hat{k}$.
It is also equivalent to the DLP when $\mathbbm{k}$ is unknown (i.e., to find $\check{t}/\hat{t} = \log_{g^{1/(\hat{k}+x)}}{g^{1/(\check{k}+x)}} = \log_{\hat{z}}{\check{z}}$).

In 2HashRSA$_N$ it requires that, for any $x$, adversaries cannot find $\hat{t}$ and $\check{t}$ satisfying that $H_2(x, x^{\hat{k}}\hat{t}/\check{t}) = H_2(x, x^{\check{k}})$ where $\check{k} \neq \hat{k}$.
When $\mathbbm{k}$ and $\{x_j^{k_i}\}$ are kept secret to adversaries,
    this property is ensured because
 adversaries cannot solve $\check{t} = \hat{t}x^{\hat{k}}/x^{\check{k}}\bmod N$ (i.e., the probability that $\check{t}/\hat{t}$ is equal to unknown $x^{\hat{k}}/x^{\check{k}}$ is negligible).
\hfill $\square$

\begin{theorem}
Strong collision resistance on 2nd-input
    is ensured in HashDH, DY$_{\textrm{HE}}$, and 2HashRSA$_N$ with secret $x_j^{k_i}$.
\label{theorem-rp-wo-checking}
\end{theorem}

\noindent\textbf{Proof.}
In HashDH,
    given known $\mathbbm{x}$, known $\mathbbm{z}$, and unknown $\mathbbm{k}$,
 if adversaries could find ${j_1}$, ${j_2}$, ${i_1}$, ${i_2}$, $t_1$ and $t_2$ which satisfy that $((x_{j_1}^{t_1})^{k_{i_1}})^{1/t_2} = x_{j_2}^{k_{i_2}}$ (i.e., $z_{i_1,j_1}^{t_1} = z_{i_2,j_2}^{t_2}$)
 where ${j_1} \neq {j_2}$,
 the DLP would be solved (i.e., to find $t_1/t_2 = \log_{z_{i_1,j_1}}{z_{i_2,j_2}}$).\footnote{As mentioned in Section \ref{system-model}, $k_i$ and $r_j$ are known to \emph{only} the honest IdP  in UppreSSO, where $ID_{U_i} = k_i$ and $ID_{RP_j}=[r_j]G$.
In the $\mathbbm{F}_q$ versions of the identity transformations,  $k_i$ and $r_j$ are kept unknown to adversaries where $x_j=g^{r_j}$;
otherwise, once $\mathbbm{k}$ or $\mathbbm{r} = \{r_{j=1, \cdots, p}\}$ is leaked, it is easy for adversaries to solve $(x_{j_1}^{t_1})^{k_{i_1}} = (x_{j_2}^{k_{i_2}})^{t_2}$ where ${j_1} \neq {j_2}$. Similar conclusions are not applicable to DY$_{\textrm{HE}}$.}
Alternatively, the adversaries may attempt to find $x'$, ${j_2}$, ${i_1}$, ${i_2}$ and $t_2$ satisfying that  $x'^{k_{i_1}/t_2} = x_{j_2}^{k_{i_2}} = z_{i_2,j_2}$ (i.e., $1/t_2 = \log_{x'^{k_{i_1}}}{z_{i_2,j_2}}$),
        but this cannot be solved due to the DLP.

In DY$_{\textrm{HE}}$
 adversaries may attempt to find ${j_1}$, ${j_2}$, ${i_1}$, ${i_2}$, $t_1$ and $t_2$ which satisfy that $(g^{1/t_1(k_{i_1}+x_{j_1})})^{t_2} = g^{1/(k_{i_2}+x_{j_2})}$, 
 i.e., $g^{t_1/t_2} = g^{(k_{i_2}+x_{j_2})/(k_{i_1}+x_{j_1})}$,
    where ${j_1} \neq {j_2}$.
This requires the adversaries to find ${t_1/t_2} = (k_{i_2}+x_{j_2})/(k_{i_1}+x_{j_1}) = \log_{z_{i_2,j_2}}{z_{i_1,j_1}}$,
    and the probability is negligible when $k_i$ is unknown.
Or, the adversaries attempt to find $x'$, ${j_2}$, ${i_1}$, ${i_2}$ and $t_2$ satisfying that  $(g^{1/Dec(x')})^{t_2} = g^{1/(k_{i_2}+x_{j_2})}  = z_{i_2,j_2}$,
        but this is also equivalent to the DLP when ${k}_i$ is unknown (i.e., to find $t_2/Dec(x') = \log_g {z_{i_2,j_2}}$).

In 2HashRSA$_N$
adversaries attempt to find (\emph{a}) ${j_1}$, ${j_2}$, ${i_1}$, ${i_2}$, $t_1$ and $t_2$ satisfying that $H_2(x_{j_2}, x_{j_1}^{k_{i_1}} t_1/t_2) = H_2(x_{j_2}, x_{j_2}^{k_{i_2}})$, 
    i.e., $x_{j_1}^{k_{i_1}} t_1/t_2 = x_{j_2}^{k_{i_2}}$,
    or (\emph{b}) $x'$, ${j_2}$, ${i_1}$, ${i_2}$ and $t_2$ satisfying that
        $H_2(x_{j_2},x'^{k_{i_1}}/t_2) = H_2(x_{j_2},x_{j_2}^{k_{i_2}})$,
        i.e., $x'^{k_{i_1}}/t_2 = x_{j_2}^{k_{i_2}}$.
When ${k}_i$ and $x_j^{k_i}$ are kept secret to adversaries,
    they cannot solve any of these problems,
    i.e., the probability is negligible that (\emph{a}) $t_1/t_2$ is equal to unknown $x_{j_2}^{k_{i_2}} / x_{j_1}^{k_{i_1}}$ or (\emph{b}) $x'^{k_{i_1}}/t_2$ is equal to unknown $x_{j_2}^{k_{i_2}}$.
\hfill $\square$
\vspace{2mm}

In a generalized UppreSSO system built based on 2HashRSA$_N$,
  $x_j^{k_i}$ is an intermediate variable of $\mathcal{UBL}()$ calculated by an RP.
Since an RP is usually protected not so well as the IdP where $k_i$ is processed internally, 
    $x_j^{k_i}$ might be leaked sometimes.
%Since an RP might be malicious can be protected not so well as the IdP where $k_i$ is processed internally, $x_j^{k_i}$ might be leaked sometimes.
If $x_j^{k_i}$ is known to adversaries, strong collision resistance on 1st-input does not hold
because
$\check{t} = x^{\hat{k}}\hat{t}/x^{\check{k}}$ satisfies that $H_2(x, \hat{z}'/\check{t}) = H_2(x, x^{\check{k}})$,
and strong collision resistance on 2nd-input does not either 
        because $t_2 = x_{j_1}^{k_{i_1}} t_1/x_{j_2}^{k_{i_2}}$ satisfies that $H_2(x_{j_2}, x_{j_1}^{k_{i_1}}t_1/t_2) = H_2(x_{j_2}, x_{j_2}^{k_{i_2}})$.\footnote{Although a malicious RP might directly share $x_j^{k_i}$ with adversaries, this does not lead to attackers because it makes no sense to consider the login results at malicious RPs.}

\begin{table*}[tb]
    \caption{Properties of Typical OPRFs}
    \centering
    \begin{tabular}{|r|c|r|c|c|c|c|} \hline
    {\textbf{OPRF Property}}  & \multicolumn{2}{c|}{{\textbf{SSO-Related Property}}} & {\textbf{HashDH}} & {\textbf{NR$_{\textrm{HE}}$}} & {\textbf{DY$_{\textrm{HE}}$}} & {\textbf{2HashRSA$_N$$^{3}$}} \\ \hline

%    \multirow{2}{*}{\textbf{OPRF Property}}  & \multicolumn{2}{c|}{\multirow{2}{*}{\textbf{SSO-Related Property}}} & \multirow{2}{*}{\textbf{HashDH}} & \multirow{2}{*}{\textbf{NR$_{\textrm{HE}}$}} & \multirow{2}{*}{\textbf{DY$_{\textrm{HE}}$}} & \multicolumn{2}{c|}{\textbf{2HashRSA$_N$}} \\ \cline{7-8}
%    & \multicolumn{2}{c|}{} &&&& Secret $x_j^{k_i}$ & Public $x_j^{k_i}$ \\ \hline

    Output uniqueness & \multicolumn{2}{c|}{{Account uniqueness}} & $\surd$ & $\bot$$^{2}$ & $\surd$ & {$\surd$} \\ \hline

    {Correctness} & \multicolumn{2}{c|}{{Account correctness}} & $\surd$ & $\surd$ & $\surd$ & {$\surd$} \\ \hline
    
    {Obliviousness} & \multirow{2}{*}{{Privacy}} & {IdP untraceability} & $\surd$ & $\surd$ & $\surd$ & {$\surd$} \\ \cline{1-1}\cline{3-7}

    {Indistinguishability of keys$^{1}$} & & {RP unlinkability} & $\surd$ & $\surd$ & $\surd$ & $\bot$ \\ \hline

    Strong collision resistance on 1st-input & \multirow{2}{*}{\makecell[c]{{Security}\\{w/o $PID_{RP}$ Checking}}} & {User identification} & $\surd$ & - & $\surd$ & $\bot$ \\ \cline{1-1}\cline{3-7}

    Strong collision resistance on 2nd-input & & {RP designation} & $\surd$ & - & $\surd$ & $\bot$ \\ \hline

    Weak collision resistance on 1st-input & \multirow{2}{*}{\makecell[c]{{Security}\\{w/ $PID_{RP}$ Checking}}} & {User identification} & $\surd$ & - & $\surd$ & $\surd$ \\ \cline{1-1}\cline{3-7}

    Weak collision resistance on 2nd-input & & {RP designation} & $\surd$ & - & $\surd$ & $\surd$ \\ \hline
\end{tabular}
    \label{oprf-property}
{
\begin{enumerate}
  \item[] \textbf{$\surd$ means that a property is ensured, and $\bot$ means not.}
  \item   Indistinguishability of keys = Pseudo-randomness + Key-identifier freeness. Some protocols are slightly revised in Section \ref{proofs-oprf},
    to ensure key-identifier freeness.
  \item   It makes no sense to discuss user identification and RP designation without account uniqueness.
%  \item[3.] Security w/o $PID_{RP}$ checking is ensured in 2HashRSA$_N$, when $x_j^{k_i}$ are kept secret to adversaries. When $x_j^{k_i}$ are publicly known in 2HashRSA$_N$, security w/o $PID_{RP}$ checking is not ensured but security w/ $PID_{RP}$ checking is ensured.
  %\item[4.] In 2HashRSA$_N$ weak key-identifier freeness is ensured but not strong key-identifier freeness, because $x'= \mathcal{BL}(x,t,\omega) = xt^e$ and $e$ identifies an OPRF key. Thus, when $x_j^{k_i}$ are publicly known and then $PID_{RP}$ is checked to ensure user identification and RP designation, RP unlinkability (or indistinguishability of keys) is broken.
  \item This column lists the properties of 2HashRSA$_N$ with pubic $x_j^{k_i}$,
   and in 2HashRSA$_N$ with secret $x_j^{k_i}$ all properties are ensured.
\end{enumerate}}
\end{table*}

We further analyze the properties of weak collision resistance in 2HashRSA$_N$ with \emph{public} $x_j^{k_i}$.
However, as generalized UppreSSO built based on an OPRF with weak collision resistance 
    requires RPs to check $PID_{RP}$ before accepting a token,
in such a system based on 2HashRSA$_N$ with public $x_j^{k_i}$,
the colluding RPs link a user by $e$ when they check $PID_{RP}$  and then RP unlinkability is broken
because $PID_{RP} = xt^e$ and $e$ uniquely identifies an OPRF key.

\begin{theorem}
Weak collision resistance on 1st-input is ensured in 2HashRSA$_N$ with public $x_j^{k_i}$,
when ${e_{i_2}-e_{i_1}} \not\in \mathbbm{e}$ for any $e_{i_2}, e_{i_1} \in \mathbbm{e}$.
\label{theorem-uid-w-checking-rsa}
\end{theorem}

\noindent\textbf{Proof.}
Given known $\mathbbm{e} = \{e_{i=1, \cdots, s}\}$, known $\mathbbm{z}$ and unknown $\mathbbm{k}$,
    for any $x$,
  adversaries attempt to find $\check{t}$, $\check{e}$ and some protocol instance $(x, \hat{x}', \hat{z}', \hat{t}, \hat{\omega})$ generated with unknown $\hat{k}$ which satisfy that 
    $x\check{t}^{\check{e}} = \hat{x}'$
    and $\hat{x}'^{\hat{k}} / \check{t} = x^{\check{k}}$,
     where $\check{k} \neq \hat{k}$ (i.e., $\check{e} \neq \hat{e}$).

These are equivalent to $x\check{t}^{\check{e}} = \hat{x}'$
    and $(\hat{x}'^{\hat{k}} / \check{t})^{\hat{e}} = \hat{x}' / \check{t}^{\hat{e}} = x^{\check{k}\hat{e}}$,
and then it requires that $\hat{x}' = x\check{t}^{\check{e}} = x^{\check{k}\hat{e}} \check{t}^{\hat{e}}$.
This requires the adversaries to solve $\check{t}$ satisfying that $\check{t}^{\check{e}-\hat{e}} = x^{\check{k}\hat{e}}/x$.
Let $\bar{e} = \check{e}-\hat{e}$,
    and only when $\bar{e} \in \mathbbm{e}$,
    $\bar{U}$ could receive $\check{t}=(x^{\check{k}\hat{e}}/x)^{\bar{k}}$ from the OPRF server by submitting $x^{\check{k}\hat{e}}/x$.

They are also equivalent to  $(x\check{t}^{\check{e}})^{\check{k}} = x^{\check{k}}\check{t} = \hat{x}'^{\check{k}}$
    and $\check{t} = \hat{x}'^{\hat{k}} / x^{\check{k}}$,
and then it requires that $\check{t} = \hat{x}'^{\check{k}} / x^{\check{k}} = \hat{x}'^{\hat{k}} / x^{\check{k}}$.
Thus, $\hat{x}' = \pm1$ or $0$.

Finally, this property is ensured, if (\emph{a}) the OPRF server (or the IdP) selects $e_i$ carefully
    to ensure ${e_{i_2}-e_{i_1}} \not\in \mathbbm{e}$ for any $e_{i_2}, e_{i_1} \in \mathbbm{e}$,
    or (\emph{b}) the IdP never signs tokens binding $x'=\pm1$ or $0$.
\hfill $\square$

\begin{theorem}
Weak collision resistance on 2nd-input is ensured in 2HashRSA$_N$ with public $x_j^{k_i}$,
when ${e_{i_2}-e_{i_1}} \not\in \mathbbm{e}$ for any $e_{i_2}, e_{i_1} \in \mathbbm{e}$.
\label{theorem-rp-w-checking-rsa}
\end{theorem}

\noindent\textbf{Proof.}
Given known $\mathbbm{x}$, known $\mathbbm{z}$,
 unknown $\mathbbm{k}$, and known $\mathbbm{e}$,
adversaries attempt to find $x'$, ${j_2}$, ${i_1}$, ${i_2}$ and $t_2$ which satisfy that
$x_{j_2}t_2^{e_{i_2}} = x'$ and $H_2(x_{j_1}, x'^{k_{i_1}} / t_2) = H_2(x_{j_2}, x_{j_2}^{k_{i_2}})$.

These are equivalent to $x_{j_2}t_2^{e_{i_2}} = x'$ and $x'^{k_{i_1}} / t_2 = x_{j_2}^{k_{i_2}}$ (i.e., $(x'^{k_{i_1}} / t_2)^{e_{i_1}} = x' / t_2^{e_{i_1}}= x_{j_2}^{k_{i_2}e_{i_1}}$),
    and then it requires that $x'= x_{j_2}t_2^{e_{i_2}} = x_{j_2}^{k_{i_2}e_{i_1}} t_2^{e_{i_1}}$.
This requires the adversaries to solve $t_2$ satisfying that $t_2^{e_{i_2}-e_{i_1}} = x_{j_2}^{k_{i_2} e_{i_1}}/x_{j_2}$.
Let $\bar{e} = {e_{i_2}-e_{i_1}}$,
    and only when $\bar{e} \in \mathbbm{e}$,
    $\bar{U}$ could receive $t_2= (x_{j_2}^{k_{i_2} e_{i_1}}/x_{j_2})^{\bar{k}}$ from the server by submitting $x_{j_2}^{k_{i_2} e_{i_1}}/x_{j_2}$.
    
They are also equivalent to $x_{j_2}t_2^{e_{i_2}} = x'$ (i.e., $(x_{j_2}t_2^{e_{i_2}})^{k_{i_2}} = x_{j_2}^{k_{i_2}} t_2 = x'^{k_{i_2}}$) and $x'^{k_{i_1}} / t_2 = x_{j_2}^{k_{i_2}}$,
    and then it requires that $t_2 = x'^{k_{i_2}} / x_{j_2}^{k_{i_2}} = x'^{k_{i_1}} / x_{j_2}^{k_{i_2}}$.
Thus, $\hat{x}' = \pm1$ or $\hat{x}' = 0$.

Alternatively, the adversaries may attempt to find ${j_1}$, ${j_2}$, ${i_1}$, ${i_2}$, $t_1$ and $t_2$ which satisfy that
     $x_{j_2}t_2^{e_{i_2}} = x_{j_1}t_1^{e_{i_1}}$ and $(x_{j_1}t_1^{e_{i_1}})^{k_{i_1}} / t_2 = x_{j_2}^{k_{i_2}}$ 
 where ${j_1} \neq {j_2}$.
Similarly, it requires that $t_2= (x_{j_2}^{k_{i_2} e_{i_1}}/x_{j_2})^{\bar{k}}$ and $x_{j_1}t_1^{e_{i_1}} = \pm1$ or $0$.

This property is ensured, if (\emph{a}) $e_i$ is selected carefully
    to ensure ${e_{i_2}-e_{i_1}} \not\in \mathbbm{e}$ for any $e_{i_2}, e_{i_1} \in \mathbbm{e}$,
    or (\emph{b}) the IdP never signs tokens binding $x'=\pm1$ or $0$.
\hfill $\square$
\vspace{2mm}

Table \ref{oprf-property} lists the properties of typical OPRF protocols.
The properties of weak collision resistance are certainly ensured, if the strong ones hold.
Thus, in addition to HashDH, the identity-transformation approach can be instantiated with DY$_{\textrm{HE}}$ and 2HashRSA$_N$.

We do \emph{not} recommend a generalized UppreSSO system built based on either 2HashRSA$_N$ or DY$_{\textrm{HE}}$.
All the properties are ensured in 2HashRSA$_N$ with secret $x_j^{k_i}$, but it requires \emph{extra} protections on $\{x_j^{k_i}\}$ by RPs.
In 2HashRSA$_N$ with public $x_j^{k_i}$, indistinguishability of keys (i.e., RP unlinkability in the UppreSSO system)
    is \emph{not} ensured.
      %  and it requires that $e_i$ is selected carefully.
Meanwhile, in a system built on DY$_{\textrm{HE}}$ it requires that $\omega = Enc(k_i)$ is \emph{never} leaked to any entity except the initiating user and the IdP;
otherwise, once $Enc(k_{i})$ is leaked in such a system where the additively HE key pair is \emph{permanent},
    an adversary could impersonate the victim user by straightforwardly using $Enc(k_{i})$ in any later login,
    which to some extend plays the same role as the credential for user authentication to the IdP.
Anyway, this attack does not contradict the proved security properties of DY$_{\textrm{HE}}$ in Theorems \ref{theorem-uid-wo-checking} and \ref{theorem-rp-wo-checking}, because in this attack the adversary actually requests an identity token on behalf of the victim user but it cannot be detected by the IdP.
    On the other hand, the security properties (i.e., user identification and RP designation) prevent a token requested by honest users from being exploited by adversaries.

\subsection{Account Synchronization}
\label{account-syn}
If new users are allowed to register at a generalized UppreSSO system after the initialization,
in order to distinguish meaningful accounts from meaningless ones,
 an RP needs to synchronize its \emph{local} account set \cite{uppresso-arxiv-all}.
% synchronizes its account list \cite{uppresso-conference} by two alternative means as below,

In generalized UppreSSO systems %built based on HashDH, DY$_{\textrm{HE}}$ or 2HashRSA$_N$ with secret $x_j^{k_i}$,
    where an RP regularly does not check $PID_{RP}$ before accepting an identity token,
    accounts are synchronized by \emph{conditional} $PID_{RP}$ checking as below.
When the RP derives some account not in its current account set,
    it \emph{additionally} checks $PID_{RP}$:
     If $PID_{RP}$ enclosed in the token is equal to $\mathcal{F}_{PID_{RP}}(ID_{RP}, t, \omega)$,
                this account is asserted to belong to a newly-registered user
                and then the RP updates its account set (i.e., the derived account is inserted into the set);
    otherwise, it is rejected.

When % the system is built based on 2HashRSA$_N$ with public $x_j^{k_i}$,
    an RP always checks $PID_{RP}$ and accepts an identity token only if $PID_{RP}$ in the token is equal to $\mathcal{F}_{PID_{RP}}(ID_{RP}, t, \omega)$,
account synchronization is automatically finished and no more operation is needed in this case.
%In fact, since an RP will not be automatically informed after $x_j^{k_i}$ is leaked, in a system based on 2HashRSA$_N$ we recommend that the RPs always check $PID_{RP}$.

\section{Related Work}
\label{related-work}

\noindent\textbf{Security and Privacy Analysis of SSO Protocols.}
Sufficient conditions of secure SSO services are presented \cite{FettKS14,BrowserID,SPRESSO}:
    (\emph{a}) An attacker cannot log into an honest RP as an account owned by any honest user, and (\emph{b})
an honest user never log into an honest RP as an account not owned by this user.
When authenticity, confidentiality and integrity of identity tokens are ensured,
        these conditions are equivalent to user identification and RP designation.
Security of the SSO services with identity transformations 
 has been proved in UppreSSO \cite{uppresso-arxiv-all},
but the variations of user identification and RP designation are comprehensively compared only in this paper.

Indistinguishability is defined to analyze user privacy (i.e., IdP untraceability) in Spresso \cite{SPRESSO},
        while privacy of generalized UppreSSO services is actually guaranteed by indistinguishability of the underlying OPRF: Private inputs are indistinguishable to the OPRF server (i.e., IdP untraceability), and OPRF keys are indistinguishable to OPRF users (i.e., RP unlinkability).

BISON \cite{bison} and OPPID \cite{OPPID} utilized the same idea to blindly transform a user identity into pseudonyms (or accounts) as UppreSSO \cite{uppresso-arxiv-all}, and pointed out that the identity transformations mathematically employ the same technique as the HashDH OPRF.
OPPID \cite{OPPID} also required the visited RP to be anonymously authenticated to the IdP before retrieving the identity tokens.
However, they do not uncover the relationship between the identity
transformations in SSO and general OPRFs, and only the proofs of HashDH are presented \cite{bison,OPPID}: The privacy properties, i.e., IdP untraceability and RP unlinkability, are proved as \emph{user-SP relationship anonymity} and \emph{pseudonym unlinkability} in BISON, respectively;
In OPPID the security properties (i.e., RP designation and user identification) are proved as \emph{session binding}, and the privacy properties are called \emph{unobservability} and \emph{unlinkability}, respectively.
%Besides, BISON requires special user devices to eliminate the assumption of honest user-i scripts, as discussed in Section \ref{arpsso-improvement}.

Dolev-Yao style models \cite{SPRESSO,BrowserID,FettKS14} are developed to analyze the communications among entities in an SSO system, to ensure that all messages including identity tokens are delivered as expected in the system
        and then to prove security and privacy of services 
        based on traditional public-key and symmetric cryptographic algorithms (e.g., RSA and AES in Spresso \cite{SPRESSO}).
On the contrary, in this paper security and privacy of SSO services are analyzed based on more complicated cryptographic primitives to transform (pseudo-)identities and accounts,
    while secure communications among entities are assumed.
   % (i.e., authenticity, confidentiality and integrity of identity tokens) 

The identity transformations assume an honest IdP, while user privacy in SSO systems with a malicious IdP is considered in Spresso \cite{SPRESSO} and ticket transparency \cite{ticket-transparency,enhanced-tt}.
This paper uncovers the relationship between the identity-transformation approach in SSO and OPRFs, while the mapping of security enhancements between SSO and X.509 certificate services has been analyzed \cite{te-sso}.

\noindent\textbf{OPRFs and OPRF-Based Applications.}
OPRF protocols \cite{strong-oprf,eff-psi-oprf,voprf-proved,oprf-ot-si,two-sided-oprf} are designed and applied for password verification \cite{opaque,pesto},
server-assisted encryption \cite{oprf-deduplication,o-kms}, % 
key recovery \cite{pp-ss},  %Memento,toppss
 computation on private inputs \cite{Catalic,strong-oprf,eff-psi-oprf,PrivateDrop}, and anonymous tokens \cite{privacypass,trusttoken},
 and the knowledge of OPRFs is well systematized \cite{sok-oprf}.

Extended properties of OPRFs are proposed in various scenarios, such as verifiability \cite{voprf-proved,Pythia-PRF,DPaSE}, 
committed inputs/outputs \cite{oprf-ot-si,pseudonym-oprf}, partial obliviousness \cite{pesto,DPaSE,Pythia-PRF}, updateability \cite{Pythia-PRF},
convertability \cite{ScrambleDB} and extendability \cite{DPaSE},
but the properties related to privacy-preserving SSO %(i.e., output uniqueness, key-identifier freeness and collision resistance on 1st/2nd-input)
    are analyzed in this paper for the first time.

\noindent\textbf{Privacy-Preserving Identity Federation.}
Identity federation enables a user registered at an IdP to be accepted by RPs.
Privacy-preserving identity federation
 prevents both IdP-based login tracing and RP-based identity linkage \cite{PseudoID,hyperledge-idemix,UnlimitID,uprov,Opaak,ELPASSO}, as (\emph{a}) no RP identity is enclosed in IdP-issued anonymous credentials and (\emph{b}) a user selects different pseudonyms to visit different RPs.

They even protect user privacy against collusive attacks by the IdP and RPs, because the pseudonyms cannot be linked when the ownership of anonymous credentials \cite{idemix,anon-credential-2001,anon-credential} is proved to RPs colluding with the IdP.
In such solutions a user holds an extra long-term secret
 to mask the relationship of the identities of visited RPs and the pseudonyms across RPs.
  %  so the services are accessed from browsers with an extension to handle this secret.
Although the term ``single sign-on (SSO)'' was also used in some schemes \cite{PseudoID, Opaak, ELPASSO, WangWS13, HanCSTW18, HanCSTWW20}, we refer to them as \emph{identity federation}.
While SSO minimizes a user's interactions with RPs,
such a user-maintained secret used to calculate pseudonyms (or accounts) in identity federation breaks the property of ``singleness'':
Once it is lost or leaked, the user has to contact all RPs to update every account derived from this secret.

\section{Conclusions}
\label{conclusion}
%Identity transformations are proposed for OIDC-compatible privacy-preserving SSO services,
%    providing both IdP untraceability and RP unlinkability.
In this paper, we investigate the identity-transformation approach of privacy-preserving SSO services
    in two aspects:
    (\emph{a}) The integration of identity transformations in OIDC,
            with suggestions to improve performance,
  and (\emph{b}) the relationship between identity transformations in SSO and general OPRFs,
    helping us to construct new identity transformations qualified for OIDC-compatible privacy-preserving SSO services.

To the best of our knowledge,
    this is the first time to uncover the relationship between identity transformations in privacy-preserving SSO services and general OPRFs,
        and prove the corresponding properties (i.e., output uniqueness, key-identifier freeness, collision resistance) of typical OPRFs,
        in addition to the basic properties of correctness, obliviousness and pseudo-randomness.
Our work significantly extends the understanding of both privacy-preserving SSO protocols and OPRFs.

%In the future, we plan to integrate more efficient mechanisms 
% (\emph{a}) for a visited RP to be anonymously authenticated in the authorization code flow of OIDC
%            and (\emph{b}) for a user to obtain $ID_{RP}$ of the visited RP.
%Meanwhile, we will study or design more OPRF schemes,
%    and analyze their SSO-related properties to build generalized UppreSSO systems.

%\newpage

% \section*{Acknowledgment}
% The authors would like to thank xxx for xxx 

\bibliographystyle{ACM-Reference-Format}
\bibliography{IDT-SSO-TC}

\end{document}